\documentstyle[12pt,epsf]{article}
\begin{document}
\language=0

\newcommand{\be}{\begin{equation}}
\newcommand{\ee}{\end{equation}}
\newcommand{\nn}{\nonumber}
\newcommand{\bv}{\begin{verbatim}}
\newcommand{\ev}{\end{verbatim}}
\newcommand{\met}{\mbox{$\rlap{\kern0.25em/}E_T$}}
\newcommand{\lsim}{\begin{array}{c}<\\[-9pt]\sim \\ \end{array}}
\renewcommand{\topfraction}{1}

\begin{titlepage}
\begin{flushright}
JINR E2-97-76 \\ hep-ph/9703251     \\
       March 1997        \\
\end{flushright}
\vspace{2cm}
\begin{center}
{\Large  \bf
Possibility of Chargino Search at LEP II}

\vspace{1cm}
{\bf A.S.Belyaev} $^{\mbox{a,}}$\footnote{e-mail:
belyaev@monet.npi.msu.su},
{\bf A.V.Gladyshev} $^{\mbox{b,c,}}$\footnote{e-mail:
gladysh@thsun1.jinr.dubna.su}

\vspace{1cm}

$^{\mbox{a}}$ {\it Skobeltsin Institute for Nuclear Physics,
Moscow State University, \\ 119 899, Moscow, RUSSIA}

\vspace{.5cm}

$^{\mbox{b}}$ {\it Bogoliubov Laboratory of Theoretical Physics,
     Joint Institute for Nuclear Research, \\
	141 980 Dubna, Moscow Region, RUSSIA}

\vspace{.5cm}

$^{\mbox{c}}$ {\it Physics Department, Moscow State University,
	119 899, Moscow, RUSSIA}

\end{center}

\vspace{1cm}

\begin{abstract}
We study the potential of the LEP200 collider for chargino production
and the possibility of the determination of some fundamental
SUSY parameters through the relation with this study. Total cross
sections for two basic signatures of pair chargino production and
for main backgrounds are calculated. The set of kinematic cuts
is proposed  for the effective background suppression and
extraction of the signal. Calculations and the MC simulation
performed give the limits on chargino and neutralino masses that
could be obtained at LEP200 with $\sqrt s =$ 200 GeV.
\end{abstract}

\end{titlepage}


\section{Introduction}

Supersymmetry~\cite{susy} is supposed to be the most promising key
for the solution of problems of the Standard Model. That is
why in recent years the search for SUSY has become one of the
most important tasks for the present and future accelerators.

According to the great variety of models with different assumptions
there are several particles which could have a mass within the
reach of present or forthcoming experiments. Among these
particles are: the lightest Higgs boson, the lightest neutralino,
and the lightest chargino.

The Higgs boson, even  if it is  discovered  at LEP II, will not give
us complete evidence that SUSY takes place. The neutralino discovery
at present is been questioned, since its production cross section
is rather small and the signature is hardly separated from
background events~\cite{grivaz}. All this makes chargino the
most promising superparticle to be discovered soon if
{\it there is} supersymmetry in the nature.

In this paper we analyse a possible chargino
detection via its creation in $e^+e^-$ collisions with subsequent
'leptonic' or 'hadronic' decay. We define these decay modes as
follows: the neutralino, neutrino, and the lepton or neutralino and
quark pair (jets) in the final state, respectively.

To study the real possibility to detect the signal from
chargino, the MC generator for the signal and background has been
created. We have also included the effects of detector resolution
and hadronization effects into our analysis. All squared matrix
elements and the most part of numerical calculations have been
made by the CompHEP package~\cite{comphep}, in which we have
implemented the part of supersymmetric standard model relevant
for our analysis.

The paper is organized as follows. In Section 2, we present a brief
description of the model we have used and discuss the theoretical and
experimental motivations for the choice of input parameters.
Then, in Section 3, we discuss possible modes of chargino
decays,  present the chargino production rate at LEP II and
perform the MC simulation and kinematical analysis of the signal and
background for different signatures with the aim to extract the
signal and suppress the back\-gro\-und. As a result of this analysis,
we designed an effective set of kinematical cuts. These cuts
allow one to extract the signal from the chargino up to the
collider kinematic limit. In Section 4, we examine the MSSM
parameters space and explore the regions which could be excluded
by the analysis of chargino search  at LEP II. Then we make some
final conclusions.

\section{The Model}

\subsection{Basic assumptions}

The framework of the present studies is the Minimal
Supersymmetric Standard Model (MSSM)~\cite{MSSM} which has
already shown itself to good advantage. It is the simplest
extension of the Standard Model (SM) that makes good use of the
idea of supersymmetry as an underlying principle.
Supersymmetry provides solutions to some of the inner problems
of SM offering a number of theoretically beautiful ways.
Also, in the context of Grand Unification idea it allows for the
real unification of fundamental interactions at the scale of
the order \\ $10^{16}$~GeV~\cite{ABF}.

Constituents of the model are the quarks and leptons of three
generations, gauge bosons and Higgs scalar fields. To these must
be added the {\it superpartners}, the particles that differ
in spin by half a unit. Thus, every gauge field of SM has its
fermionic superpartner, and every matter field has a scalar
partner. Contrary to SM, one needs an additional doublet of
Higgs scalar fields to give masses to up quarks and to down
quarks and leptons and to avoid the gauge anomaly.

An appropriate mathematical language to describe a supersymmetric
model is the language of superfields. The Yukawa interactions are
defined by the superpotential that in the case of MSSM reads
\be
W= \epsilon_{ij} \left( h_E L^j E^c H_1^i + h_D Q^j D^c H_1^i +
h_U Q^j U^c H_2^i + \mu H_1^i H_2^j \right)
\end{equation}
Here $Q$ and $L$ are the left-handed quarks and leptons
superfield doublets; $U^c,D^c,E^c$, superfields corresponding
to the right-handed quarks and leptons; and $H_{1,2}$ are
the Higgs superfields; $i$ and $j$ are the $SU(2)$ indices
($\epsilon_{12}=1$), color and flavor indices being understood.

Since we do not observe the exact supersymmetry in the nature,
{\it i.e.} there are no pairs of particles we could identify as
superpartners, it is a broken symmetry. At present {\it there is}
a phenomenologically acceptable way to break supersymmetry,
to include the soft breaking terms into the SUSY lagrangian.
{\it Soft}, in this context, means that these terms do not
introduce new quadratic divergences into the theory. It is
assumed that breaking of supersymmetry takes place in the hidden
sector which interacts with the visible world only via gravity.
All the possible soft terms have been studied in
Ref.~\cite{soft}. The SUSY breaking terms in MSSM are the mass
terms for the scalars, the mass terms for the gauginos and
Yukawa type terms:
\begin{eqnarray}
- L_{SB} &=& m_0^2 \sum\limits_i |\phi_i|^2 +
   \Bigl( \frac12 m_{1/2} \sum\limits_\alpha
   \lambda_\alpha \lambda_\alpha   \\
   && + A \left( h_E l^j e^c h_1^i + h_D q^j d^c h_1^i +
   h_U q^j u^c h_2^i \right) + B \mu h_1^i h_2^j + h.c. \Bigr)
   \nn
\end{eqnarray}

In general, MSSM contains too many new unknown parameters and.
To reduce the number of them, one usually makes a number of
simplifying assumptions. Some of them come from the Grand
Unified Theories and/or Supergravity theories. The most often
used are  the gauge couplings unification and the universality of
the soft supersymmetry breaking terms at the GUT scale, and
radiative bre\-ak\-ing of the $SU(2) \times U(1)$ symmetry at the
electroweak scale. We assume also the $R$-parity conservation,
which means, in particular, that su\-per\-par\-ti\-cles can be produced
only in pairs, and there exists the {\it stable} lightest
supersymmetric particle (LSP) which is usually considered the
lightest neutralino.

After the above-mentioned assumptions are made only five new
parameters are left:
$$
m_0,\; m_{1/2},\; \mu,\; A,\; \tan\beta,
$$
where
$m_0$ is a common mass for scalars at the unification scale,
$m_{1/2}$ is the same for fermions,
$\mu$ is the Higgs mixing parameter,
$A$ is the soft supersymmetry breaking parameter and
$\tan\beta$ is the ratio of vacuum expectation values of the
two Higgs fields $\tan\beta = v_2/v_1$.

Fermionic partners of the electroweak gauge and Higgs bosons
(gau\-gi\-nos and higgsinos) mix to give the mass eigenstates called
the char\-gi\-no and neutralino.

Masses can be obtained by the diagonalization of the mass
matrices:
\be
M^{(c)}=\left(
\begin{array}{cc}
M_2 & \sqrt2 M_W \sin\beta \\  \sqrt2 M_W \cos\beta & \mu
\end{array} \right)
\ee

\be
M^{(0)}=\left(
\begin{array}{cccc}
M_1 & 0 & - M_Z \cos\beta \sin\theta_W &
M_Z \sin\beta \sin\theta_W \\
0 & M_2 & M_Z \cos\beta \cos\theta_W &
- M_Z \sin\beta \cos\theta_W \\
- M_Z \cos\beta \sin\theta_W &
M_Z \cos\beta \cos\theta_W &
0 & -\mu \\
M_Z \sin\beta \sin\theta_W &
- M_Z \sin\beta \cos\theta_W &
-\mu & 0
\end{array} \right)
\ee

$M_2$ is the mass of the wino, whereas $M_1$ is the mass of the
$U(1)_Y$ gaugino, the bino.

The eigenvalues corresponding to the chargino masses are
{
\begin{eqnarray*}
\tilde m^2_{\chi_{1,2}^{\pm}}&=&\frac{1}{2}\left[M_2^2+
\mu^2+2M_W^2\right. \\
&&\left.\pm \sqrt{(M_2^2-\mu^2)^2+4M_W^4\cos^2 2\beta
+4M_W^2(M_2^2+\mu^2+ 2M_2\mu\sin 2\beta )}\right].
\end{eqnarray*}
}

The neutralino masses $\tilde m_{\chi^0}$ are roots
$\lambda _k$ of the quartic equation $F(\lambda )=0$, where
\begin{eqnarray}
F(\lambda )&=&\lambda ^4-(M_1+M_2)\lambda ^3-
(M_Z^2+\mu ^2-M_1M_2) \lambda ^2  \nonumber \\
&&+[(M_2\sin^2\theta _W+M_1\cos^2\theta _W-\mu\sin 2\beta )
M_Z^2+(M_1+M_2)\mu^2]\lambda  \nonumber \\
&&+[(M_2\sin^2\theta _W+M_1\cos^2\theta _W)\mu M_Z^2
\sin 2\beta -M_1M_2\mu^2]
\end{eqnarray}

The diagonalization can be performed in a
straightforward way by multiplying the mass matrices by
unitary rotating matrices:
\be
U^T M^{(c)} V = M^{(c)}_{diag}, \;\;\;\;\;
N^T M^{(0)} N = M^{(0)}_{diag}
\ee

\be
U = O_{-}, \;\;\;
V = \left\{ \begin{array}{ll}
O_{+}, & \mbox{det} M^{(c)} \ge 0 \\
\sigma_3 O_{+}, & \mbox{det} M^{(c)} < 0 \\
\end{array} \right. , \;\;\;
O_{\pm} =
\left( \begin{array}{rl}
\cos\phi_\pm & \sin\phi_\pm \\
-\sin\phi_\pm & \cos\phi_\pm \\
\end{array} \right),
\ee
and the angles $\phi_\pm$ are defined by
\be
\tan 2\phi_- = 2\sqrt2
\frac{\mu\sin\beta + M_2\cos\beta}{M_2^2-\mu^2-2M_W^2\cos 2\beta}
\ee
\be
\tan 2\phi_+ = 2\sqrt2
\frac{\mu\cos\beta + M_2\sin\beta}{M_2^2-\mu^2-2M_W^2\cos 2\beta}
\ee

The diagonalizing matrices $U$, $V$, and $N$ enter into the Feynman
rules.

\subsection{Model parameters}

In this subsection we describe the constraints imposed on the
model parameters and the  experimental limits on masses of
SUSY particles\footnote{The supersymmetric Standard Model
with parameters restricted by experimental constraints is often
referred to as the Constrained Minimal Supersymmetric Standard
Model.}.

Since we are dealing with the chargino and neutralino, not all the
parameters mentioned in the above discussion are relevant for
our study. As has been mentioned in {\it e.g.}~\cite{param}, the
physical properties of the chargino and neutralino depend on
$\mu$, $m_{1/2}$ and $\tan\beta$, which can be easily seen from
the mass matrices.

Numerical values of the parameters should obey some common
restrictions. For example, if one wants supersymmetry to solve
the hierarchy problem, the masses of superpartners have to be
below 1~TeV. This leads to some obvious boundaries in the
parameter space. Usually it is supposed that
$$
- 1000 \; GeV \lsim \mu \lsim 1000 \; GeV, $$
\be 
0 \lsim m_{1/2} \lsim 1000 \; GeV, \;\;\;\;\;
1 \le \tan\beta \le 50
\ee

Among the experimental constraints that can be imposed on the
model parameters the {\it top-mass constraint} deserves special
comment. It has been shown in Ref.~\cite{BEK2} that once it is
taken into account ($m_t = 175 \pm 6$ GeV) one has to distinguish
between two possible scenarios determined by the value of
$\tan\beta$. The two allowed regions are $1 < \tan\beta < 3$
and $20 < \tan\beta < 40$.
The sign of the Higgs mixing parameter $\mu$ is undetermined
since the {\it electroweak symmetry breaking constraint}
determines only $\mu^2$. One should note also that if the  {\it
relic density constraint} is included, only the high  $\tan\beta$
scenario permits the light chargino and neutralino, which is
considered the best candidate for the  dark matter of the
Universe.  In our study we used, as input parameters, the results
of the global fit analysis~\cite{BEK2,BEK3};  the latter
constraint has not been considered in Ref.\cite{BEK2},thus giving
the possibility for the light chargino and neutralino for the
low $\tan\beta$ scenario.  From now on we shall deal with both
scenarios, keeping in mind that they imply {\it different} sets
of the basic parameters with the allowed lower mass of the chargino.

Numerical values of the parameters are presented in
table~\ref{fit} for scenarios with low and high $\tan\beta$.
To calculate the mass spectrum of superparticles we run
one-loop renormalization group equations from the unification
point down to the low energy ($\approx 1$~TeV) region.
The mass spectrum obtained is shown in table~\ref{mass} for both
scenarios.

\section{Chargino production and decay}

All analytic and numerical calculations of a $2 \to 2$ process of pair
chargino production ($e^+e^- \to \chi^+\chi^-$), $2 \to 3$ processes of
chargino decay, and $2 \to 4$ background processes  
have been done with the aid of the CompHEP software
package~\cite{comphep}. This package allows one to perform complete
tree level calculations in the framework of any fed model.
For calculations in the framework of MSSM the necessary part of
the model related to the process under study has been implemented
into CompHEP. We used the Feynman rules written down in
Ref.~\cite{rosiek}. The model was extensively tested. One of the
tests was calculations both in the t'Hooft-Feynman and unitarity gauges
which had an agreement at the level of numerical accuracy.

\begin{figure}[h]
  \begin{center}
    \leavevmode
    \epsfxsize=15cm
    \epsffile{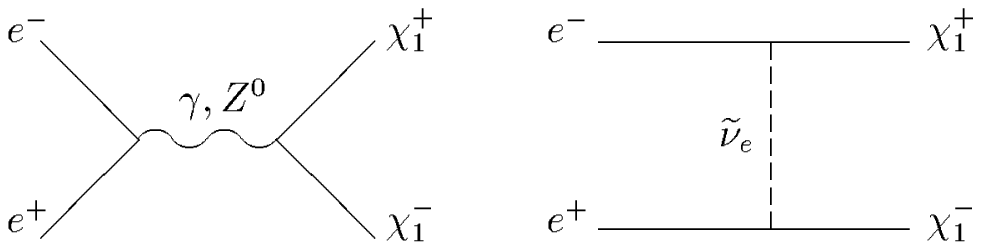}
    \vspace*{-17.0cm}
    \caption{\label{sig-diag} Diagrams for the pair chargino production
     in $e^+ e^-$ collisions.}
  \end{center}
\end{figure}

\begin{figure}[ht]
  \vspace*{-1.0cm}
  \begin{center}
    \leavevmode
    \epsfxsize=12cm
    \epsffile{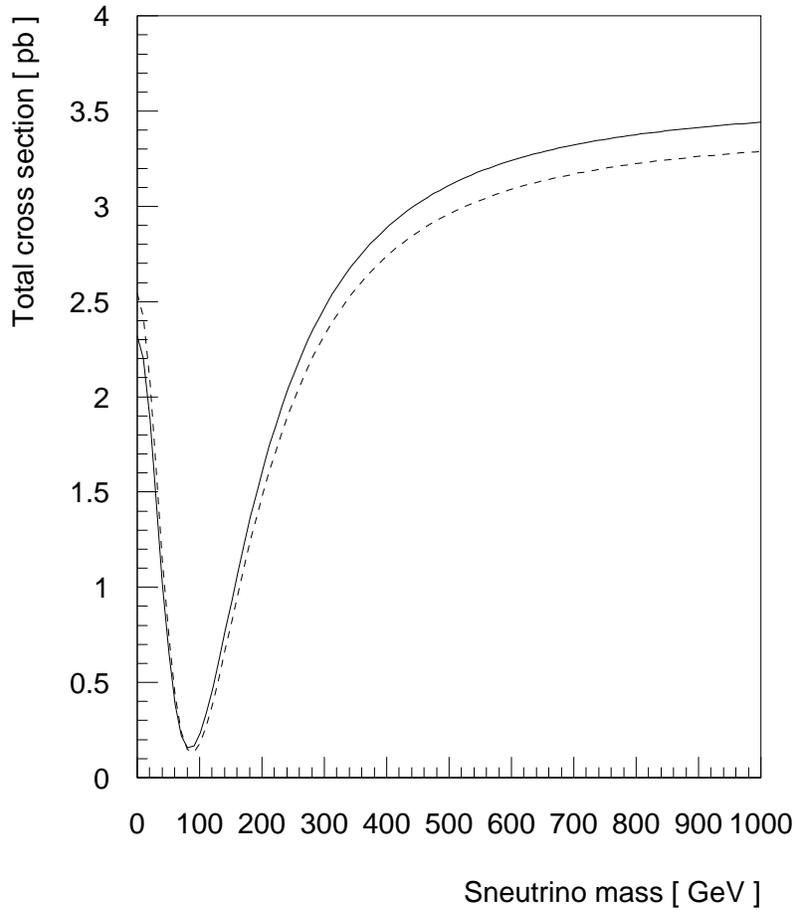}
    \caption{\label{sig-mne} The cross section as a function of the
	sneutrino mass for the low $\tan\beta$ (solid line) and high
	$\tan\beta$ scenarios ($m_{\chi^\pm}$=95 GeV).}
  \end{center}
\end{figure}

The chargino is produced  via $s$-channel (with $\gamma$ and $Z$-boson
exchange) as well as via the $t$-channel diagram with the sneutrino
illustrated in fig.~\ref{sig-diag}. $s$-channel and
$t$-channel diagrams interfere destructively, so that the cross section
has a minimum at a value of the sneutrino mass around $m_{\chi^+}$, see
fig.~\ref{sig-mne}.

For the mass spectrum under study the sneutrino mass is much higher than
that of the chargino and thus the $t$-channel diagram has a several
per cent negative  contribution. It is clearly illustrated in
fig.~\ref{sig-mne}, where the dependence of the total cross section of
pair chargino production is shown as a function of the sneutrino mass
for low $\tan\beta$ (solid line) and high $\tan\beta$ (dashed line)
scenarios.  For example, for a given sneutrino mass (407 GeV and 801
GeV) and 95 GeV chargino, cross sections are: 2.9~pb and 3.2~pb
respectively, while contribution only from the $s$-channel diagram is
3.5~pb and 3.4~pb for these two scenarios.

\begin{figure}[t]
  \vspace*{-1.0cm}
  \begin{center}
    \leavevmode
    \epsfxsize=12cm
    \epsffile{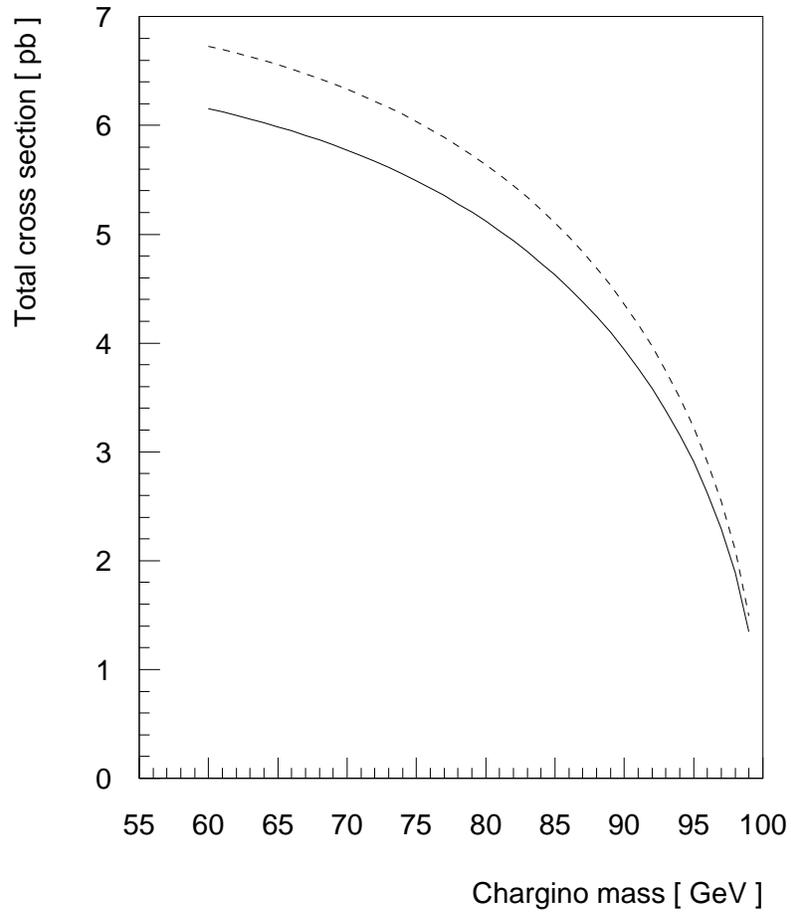}
    \caption{\label{sig-mass} The cross section as a function of
	the chargino mass for the low $\tan\beta$ (solid line) and
	high $\tan\beta$ scenarios.}
  \end{center}
\end{figure}

For the low $\tan\beta$ scenario with a lower sneutrino mass, the
negative interference is bigger than that for the high $\tan\beta$
scenario (20\% in comparison with 6\%). In the range of chargino mass
60-100 GeV the cross section is of order of several~pb varying from
6.2~pb to 1.3~pb and from 6.7~pb to 1.5~pb for the low and high
$\tan\beta$, respectiely.  The total cross section
versus chargino mass is presented in fig.~\ref{sig-mass} for both these
cases.

\begin{figure}[tb]
  \vspace*{-2.0cm}
  \begin{center}
    \leavevmode
    \epsfxsize=15cm
    \epsffile{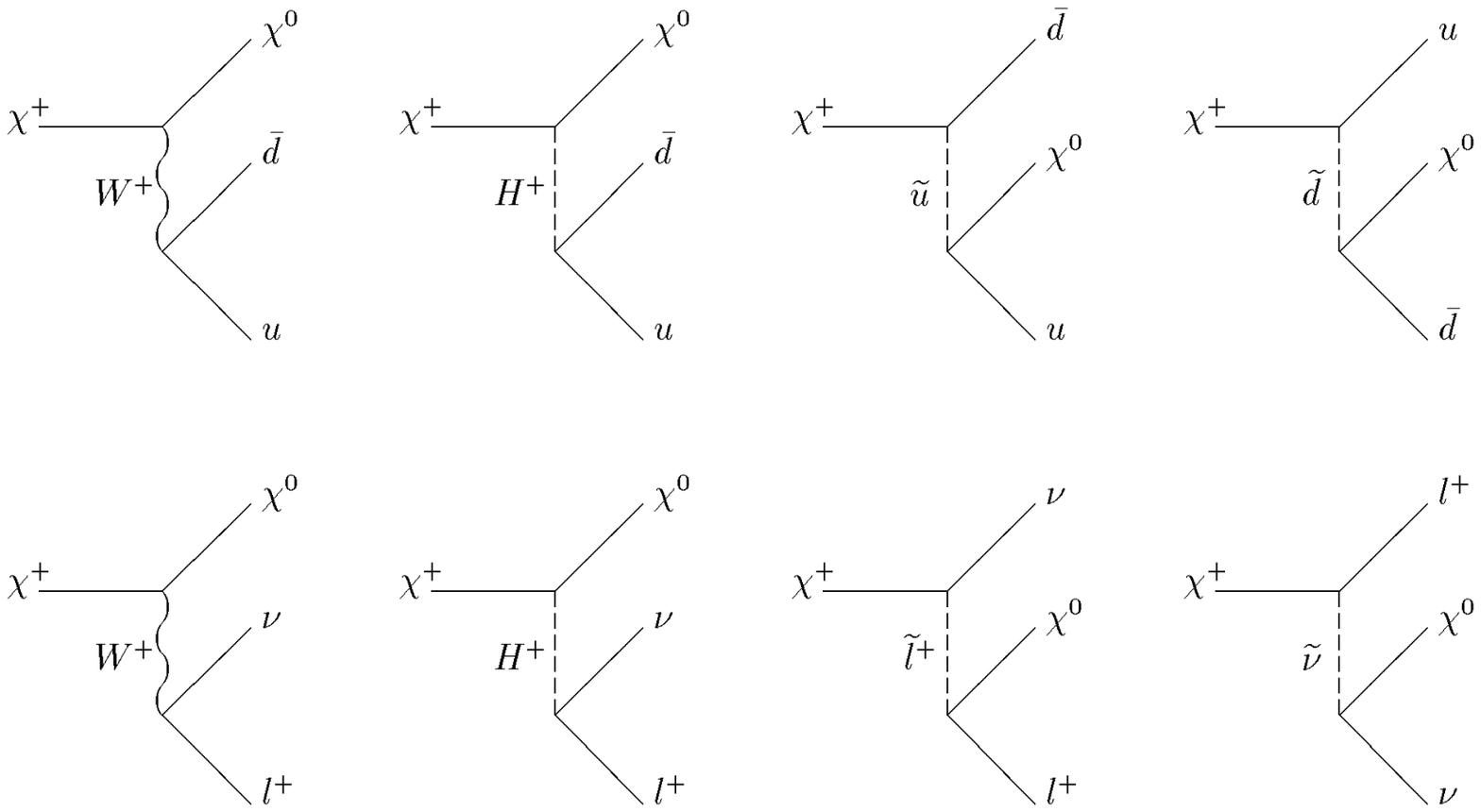}
    \vspace*{-13.0cm}
    \caption{\label{decay-diag} Diagrams for the chargino decay modes.}
  \end{center}
\end{figure}

Since the sum of neutralino and $W$-boson masses is higher than
the chargino mass, there is only the possibility of three-body decay
for the chargino. The chargino decays into
$jj+\chi^0$ or  $l+\nu + \chi^0$. The complete gauge invariant set
of Feynman diagrams is shown in fig.~\ref{decay-diag}.

If the mass of the next-to-lightest neutralino $\chi^0_2$ is lower
than the chargino one, then it could
open the cascade decay of $\chi^0_2$. But the difference between
masses of  $\chi^0_1$ and $\chi^0_2$ of several
dozen~GeV (for example, 30~GeV for the high $\tan\beta$ scenario)
results in that the branching ratio of chargino
decay into  $\chi^0_1+f+f'$  is  about 50 times
as high as  that of  decay  $\chi^0_2+f+f'$ into
(for example, 40 times for
$\chi^0_2+f+f'$  for the 95~GeV chargino, $m_{\chi^0_1}= 40$~GeV and
$m_{\chi^0_2}=70$~GeV). In the  the following we do not take the latter
decay channel into account.

The main contribution to the total decay width comes from the diagram
with the virtual $W$-boson, while the contribution from
diagrams with a heavy (300-400~GeV) selectron,
sneutrino and charged Higgs is fairly small (1-2\%).
In Table~\ref{width} we present the decay widths for the 95~GeV chargino
with respective branching ratios.

\begin{figure}[t]
  \vspace*{-1.0cm}
  \begin{center}
    \leavevmode
    \epsfxsize=12cm
    \epsffile{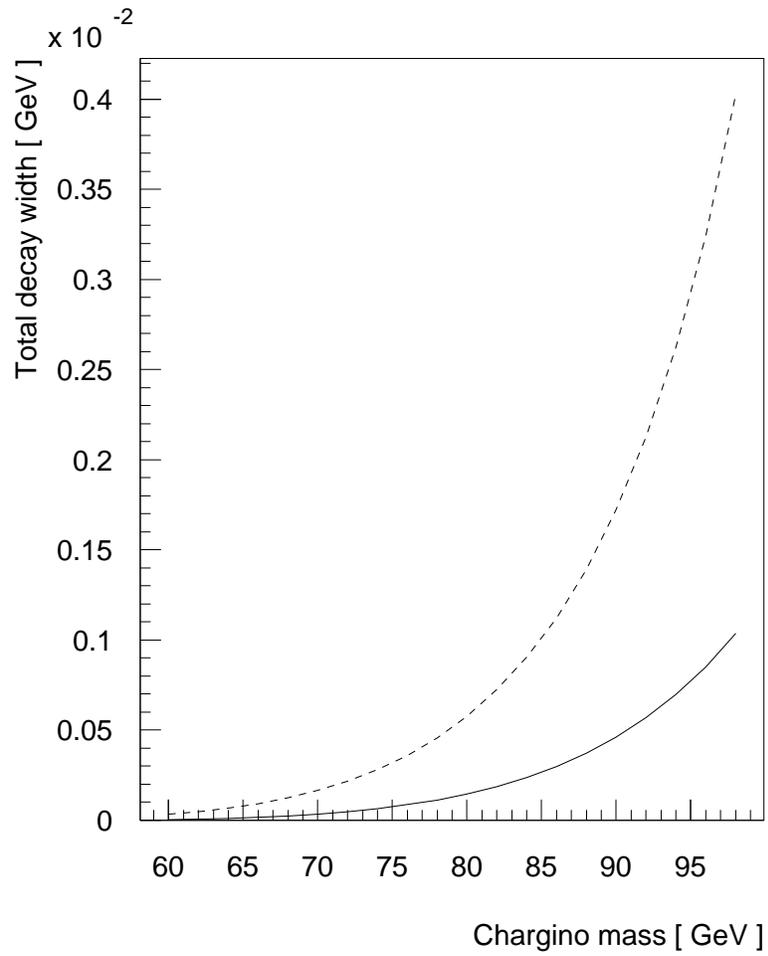}
    \caption{\label{width-mass} The total decay width as a function of
     the chargino mass for the low $\tan\beta$ (solid line) and high
	$\tan\beta$ scenarios.}
    \end{center}
\end{figure}

Though the chargino decay width is small, it has rather short lifetime
in the mass range
70-100~GeV to decay inside the detector.
The decay width as a function of the chargino mass is shown in
fig.~\ref{width-mass}.

As one can see, $jj+\chi^0$ or $l+\nu + \chi^0$ branching ratios
are very close to jet or lepton branching ratios of the $W$-boson. This
is due to the fact that the main contribution to decay widths comes
from the diagram with the virtual $W$-boson decaying into two
jets or  the lepton and neutrino.  It has been checked that these
branching ratios are almost independent of the chargino mass in the
interval 70-100~GeV.

The possible signatures for pair chargino production are:
\begin{itemize}
\item[1)] two leptons (an electron or a muon) + missing $P_T$ if two
charginos decay leptonically;
\item[2)] a charged lepton, two jets and a missing transverse momentum
in the final state if one of the charginos decays leptonically while
the other has the hadronic decay mode;
\item[3)] four jets and a missing
transverse momentum if both charginos decay hadronically.
\end{itemize}

All these signatures have the same
source of background, namely,
the $WW$ one. Pure leptonic  signature has the lowest rate. Moreover,
two neutrinos in the final state with large $P_T$ cause the problem
of reconstructing basic specific  kinematic characteristics which
are important for the signal and background separation. The last two
signatures seem to be  the most promising for chargino search.
In this paper we have concentrated just on these two cases.

For further calculations and the MC simulation we use the chargino mass
equal to 95~GeV which is almost  at the limit of the maximum
expected beam energy.

\section{Signal and Background Study}

\subsection{MC simulation}

To study the possibility of the signal extraction
from the background, the MC generator for chargino pair production and
decay  has been created. It was designed as a new user  process for
PYTHIA~5.7/JETSET~7.4  package~\cite{pythia}.
This generator is related with PYTHIA~5.7 by a
special interface and uses FORTRAN codes for squared matrix elements
produced by by CompHEP.  For integration of the squared matrix element
over the phase space and a consequent event simulation the Monte-Carlo
generator uses the BASES/SPRING package~\cite{bases}.  In the same
manner the generator for the real $2 \to 4 \;$ : $ e^+e^- \to l+jj+\nu$
background process has also been created.

The effects of the final state radiation, hadronization and string
fragmentation (by means of JETSET~7.4) have also been taken into
account.  For the detector simulation the LUCELL subroutine has been
used.  The following resolutions which are currently available at the
ALEPH detector at LEP have been used for the jet and electron energy
smearing:  $\sigma^{hadron}/E=0.8/\sqrt{E}$ and
$\sigma^{electron}/E=0.19/\sqrt{E}$~\cite{aleph}. In our analysis
we used the cone 0.4
$(\Delta R= \sqrt{\Delta\varphi^2 + \Delta\eta^2}=0.4)$
algorithm for the jet reconstruction. The minimum $E_T$ threshold for a
cell to be considered as a jet initiator has been chosen 2~GeV while
the one of summed $E_T$ for a collection of cells to be accepted as a
jet has been chosen 9~GeV.

Under the assumptions mentioned above the kinematic features of both
signatures for signal and background have been studied.

\subsection{$l+jj+\met$ signature}

Let us consider the 'mixed' mode of chargino  decay with a charged
lepton and two jets in the final state. The  branching ratio for
the chargino decay when one chargino decays into an electron or a muon;
and the other one, into two jets is: $BR(\chi^{pm} \to ljj+\met)=
2\cdot(0.658\cdot0.228)=0.30$.  For the 95~GeV chargino the cross
section of this particular channel is equal to:
$\sigma(\chi^{pm} \to ljj+\met) =
\sigma_{tot} \cdot BR(\chi^{pm} \to ljj+\met) \cdot 2.9 \cdot 0.3 =
0.9$~pb.

The main background  with the same  signature is the
$e^+e^- \to W^+W^-$ process if one of the $W$'s decays leptonically
and the other one decays hadronically. The total cross section of this
process at the tree level for $\sqrt{s}=200$~GeV is
$\sigma=20.24$~pb with the branching ratio of the $W$-bosons decay
$BR(W^{pm} \to ljj+\met)=2\cdot(0.667\cdot0.222)=0.29$.

\begin{figure}[t]
  \vspace*{-1.0cm}
  \begin{center}
    \leavevmode
    \epsfxsize=13cm
    \epsffile{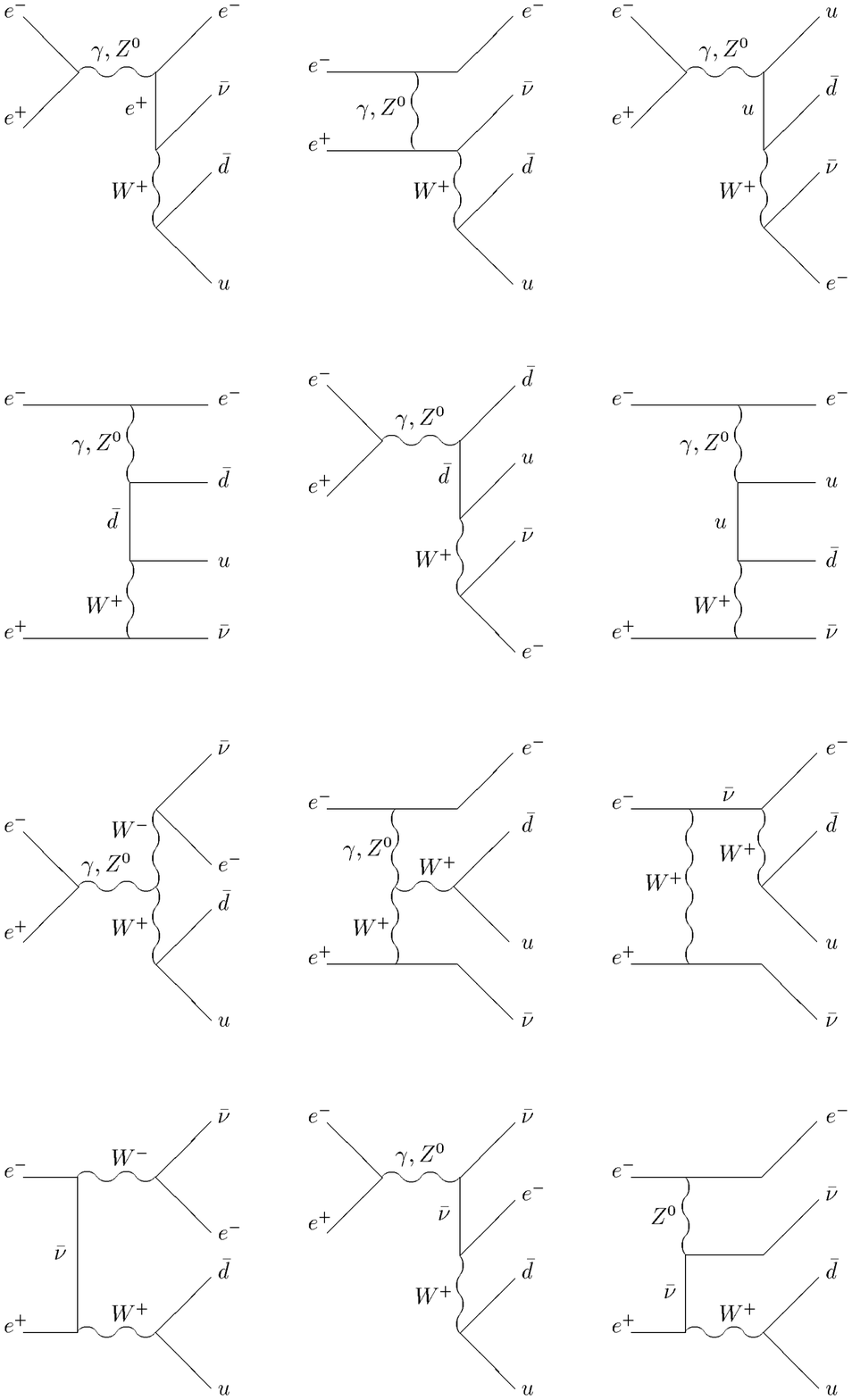}
    \vspace*{-1.0cm}
    \caption{\label{ee-enud} The complete set of diagrams for the
    $e^+e^-\to e^-\nu\bar{d}u$ process}
  \end{center}
\end{figure}

We have checked that the real $2 \to 4 \;\;\; e^+e^- \to l+jj+\nu$
process gives additional $13\%$ to the contribution from
$e^+e^- \to W^+W^-$, because of other additional diagrams.
In fig.~\ref{ee-enud}  we present, as an example, the complete set
of Feynman diagrams for $e^+e^- \to e + \nu + u + \bar{d}$  process.
Its cross section is equal to 0.78~pb, while the
contribution from $e^+e^- \to W^+W^-$ is equal to 0.69~pb. It has
been checked that the  real $2 \to 4 \;\;\; e^+e^- \to l+jj+\nu$ process
has a negligible difference in the shape distribution of main
kinematic variables in comparison with $e^+e^- \to W^+W^-$
(in $P_T$ of jets, missing $P_T$, the invariant di-jet and the
electron-neutrino mass). Thus one can use the resonance
$e^+e^- \to W^+W^-$ process with applied factor 1.13.

The total cross section of the $WW$ background for this signature is
expected to be equal to $20.24\cdot0.29\cdot1.13=6.6$~pb.

If we take into account the integral luminosity as high as
1000~pb$^{-1}$, then in terms of the number of events one can expect 900
and 6600 events, respectively, for these particular signatures.

\begin{figure}[t]
  \vspace*{-1.0cm}
  \begin{center}
    \leavevmode
    \epsfxsize=12cm
    \epsffile{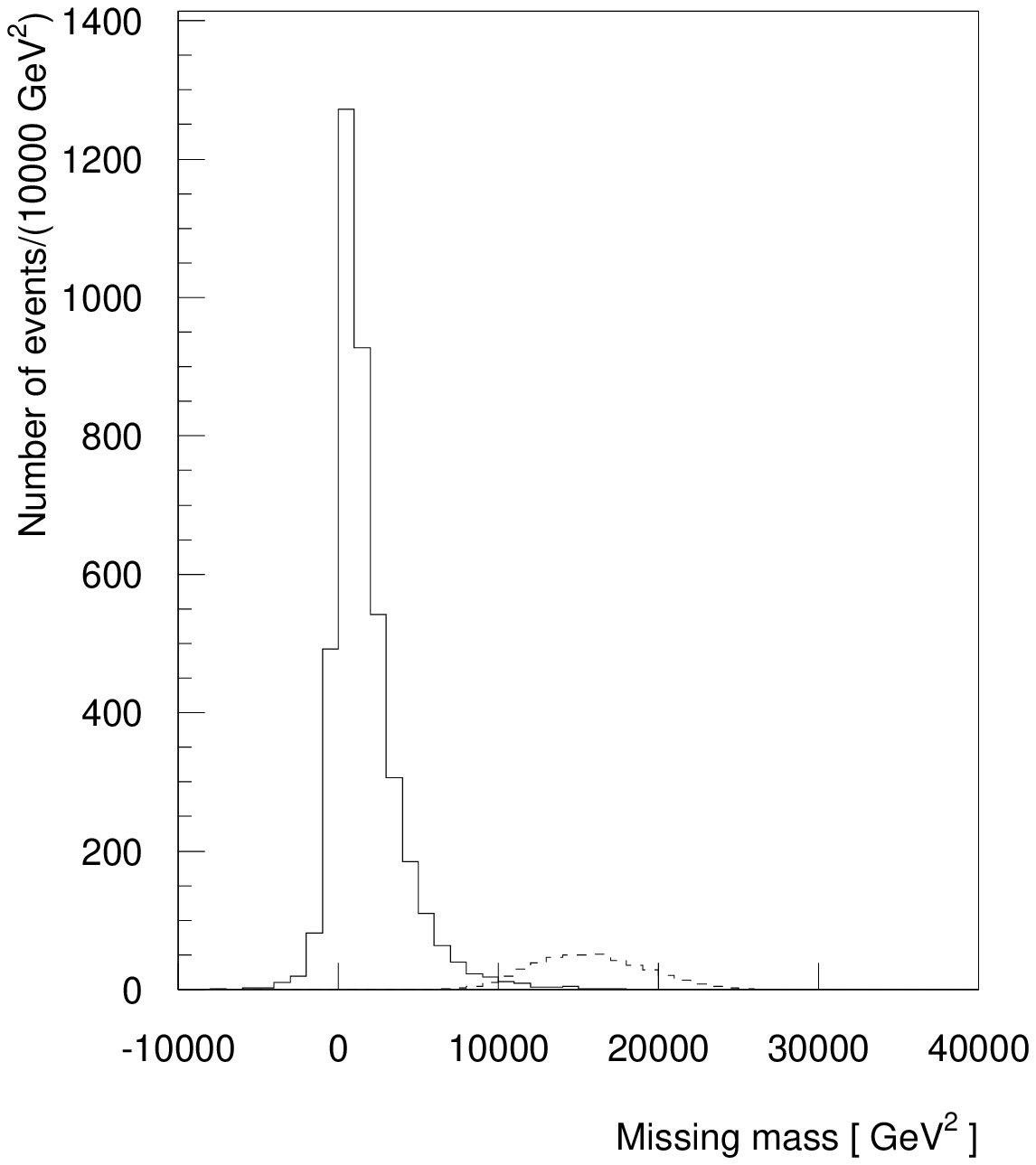}
    \caption{\label{miss-mass} The missing mass distribution
     for the $lepton+2jets+\met$ signature for the background
    (solid line) and for the chargino pair production
    (dashed line).}
  \end{center}
\end{figure}

\begin{figure}[t]
  \vspace*{-1.0cm}
  \begin{center}
    \leavevmode
    \epsfxsize=12cm
    \epsffile{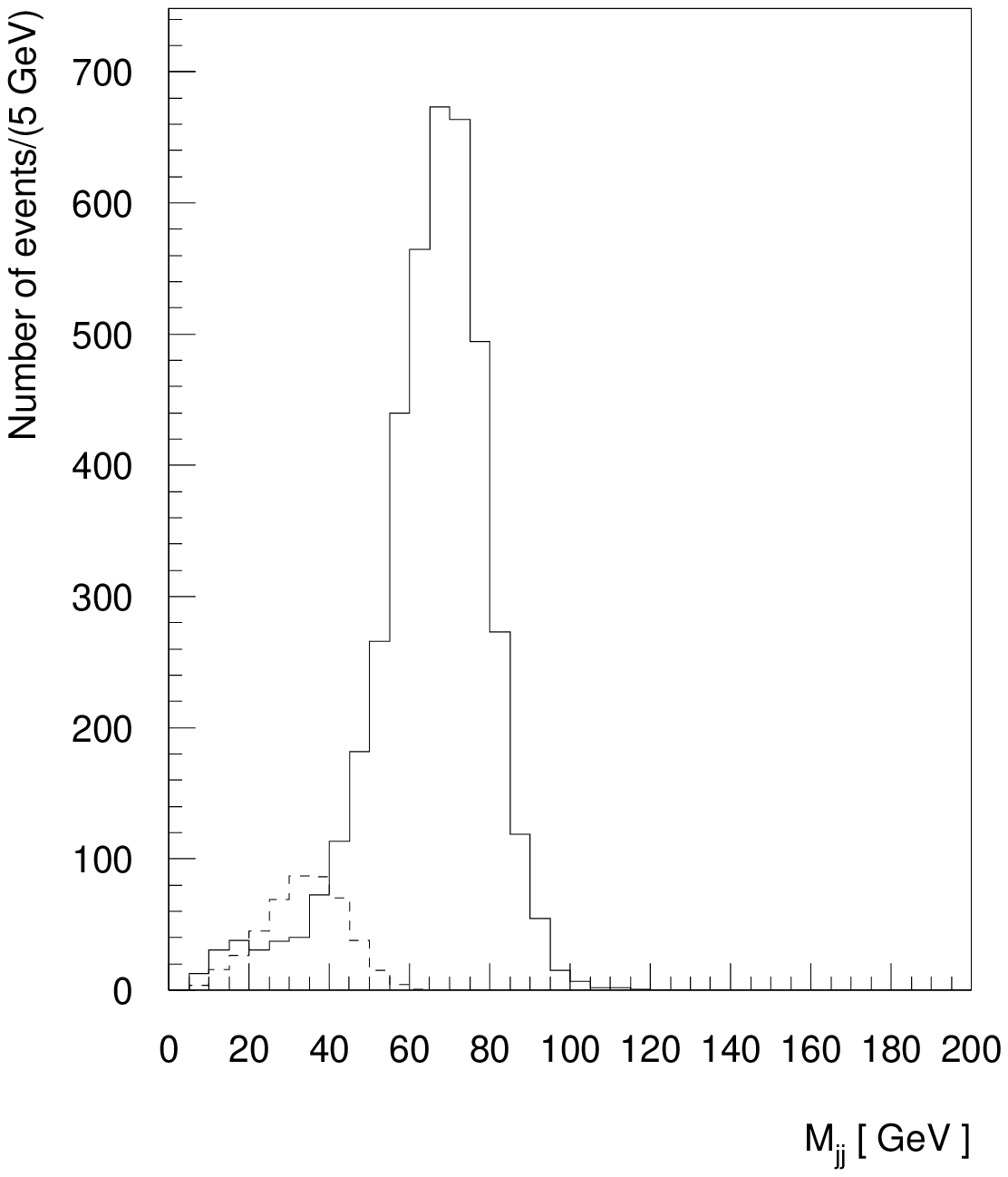}
    \caption{\label{dijet-mass} The di-jet mass distribution
     for the $lepton+2jets+\met$ signature for the background
	(solid line) and for the chargino pair production
     (dashed line).}
  \end{center}
\end{figure}

\begin{figure}[t]
  \vspace*{-1.0cm}
  \begin{center}
    \leavevmode
    \epsfxsize=12cm
    \epsffile{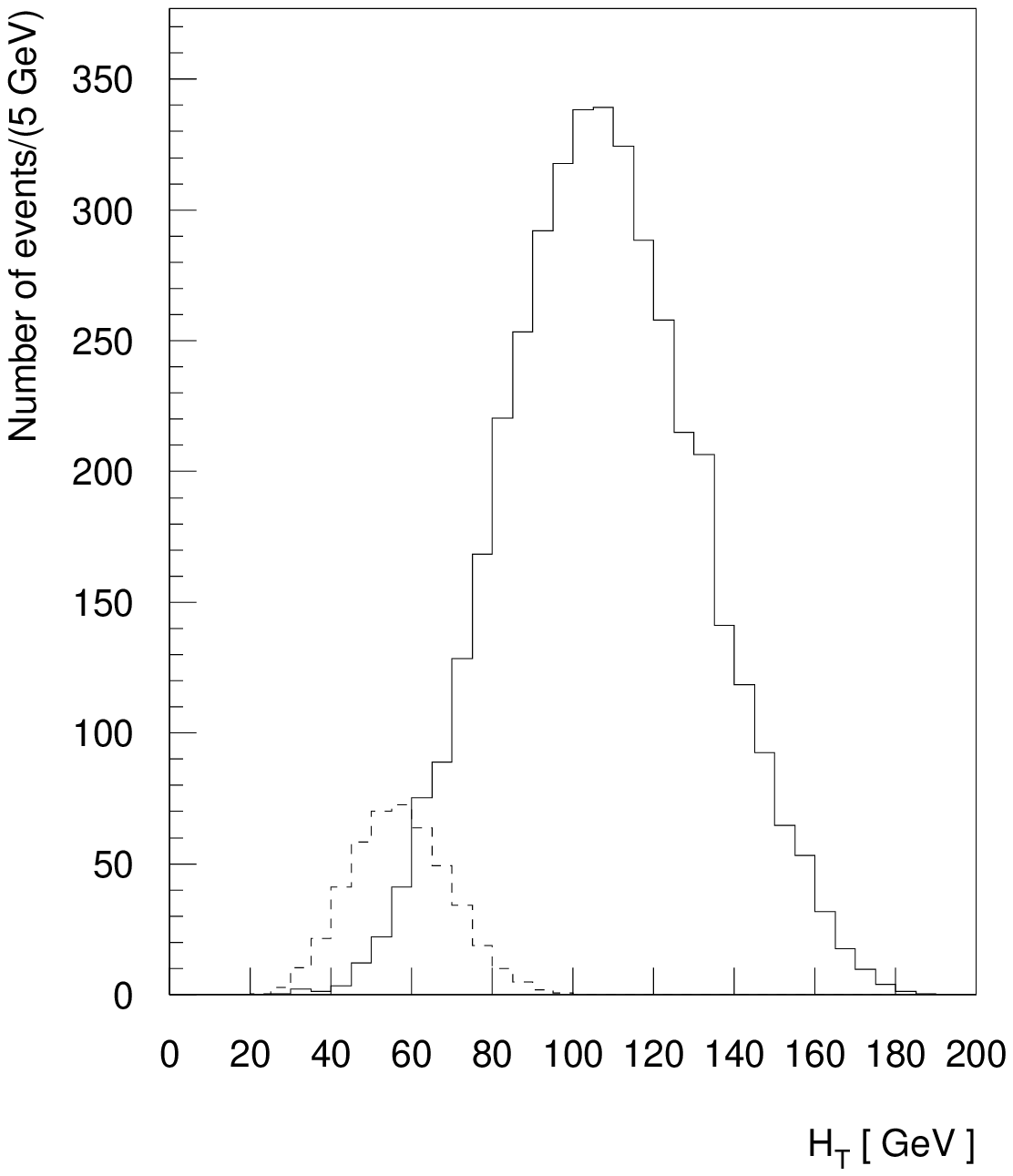}
    \caption{\label{ht} The $H_T$ distribution
     for the $lepton+2jets+\met$ signature for the background
	(solid line) and for the chargino pair production
     (dashed line).}
  \end{center}
\end{figure}

Among the kinematic variables for separation of the signal and
background which have been studied in several  papers (see, {\it e.g.}
\cite{grivaz,feng}) the most attractive are:
\begin{itemize}
\item[1)] The squared missing mass defined as missing
4-momentum squ\-a\-red.
\mbox{4-momenta} can be resolved for $e^+e^-$ because we know exactly
the energies of colliding beams and momenta of outgoing detectable
particles. For the background the squared missing mass has a peak
around zero while for the signal with undetectable heavy neutralinos it
has a peak at a value larger than $2m_{\chi^0}^2$. This fact is clearly
demonstrated in fig.~\ref{miss-mass}. This figure clarifies also the
importance of hadronization and fragmentation effects which should
be taken into account as well as the smoothing ofjet and electron
energies.  All these effects lead to the smoothing of the missing mass
distribution for the background around zero which is exactly zero at
the parton level.
\item[2)] The invariant di-jet mass which for the
$WW$ background is concentrated around the $W$ mass, while for the
signal it has a peak around $m_{\chi^+}-m_{\chi^0}$, see
fig.~\ref{dijet-mass}. As we can see, because of the jet reconstruction
effects and energy smearing, the di-jet mass distribution for the
background is shifted by approximately 10~GeV from the $Z$-peak to a
lower value
\item[3)] The invariant electron-neutrino mass, which for
the $WW$ background is concentrated around the $W$ mass, but has a peak
at values higher than the $W$-mass because of the errors in
identification of the neutrino momentum. The signal distribution for the
electron-"neutrino" mass has a peak, of course, at  much higher values
(at 155~GeV for the 95~GeV chargino and 40~GeV neutralino).
\item[4)] $H_T$ variable defined as a scalar sum of the
transverse energies of the two final state quarks and $E_T$ of the
charged lepton:  \be H_T = |E_T({\rm{jet1}})| + |E_T({\rm{jet2}})| +
	|E_T({\rm{lepton}})|.
\ee
This variable which was not considered in the previous studies  of
SUSY particle searches at LEP, but was successfully applied to the
top-quark discovery at the Tevatron~\cite{top}. It can be considered as
a measure  of massiveness of the final state particle. In our case
$H_T$ is softer for the signal than for the $WW$ background because of
a large mass which is gone away with two massive neutralinos. It is
also a strong discriminant between the signal and background, see
fig.~\ref{ht}.
\end{itemize}

\begin{figure}[t]
  \vspace*{-1.0cm}
  \begin{center}
    \leavevmode
    \epsfxsize=12cm
    \epsffile{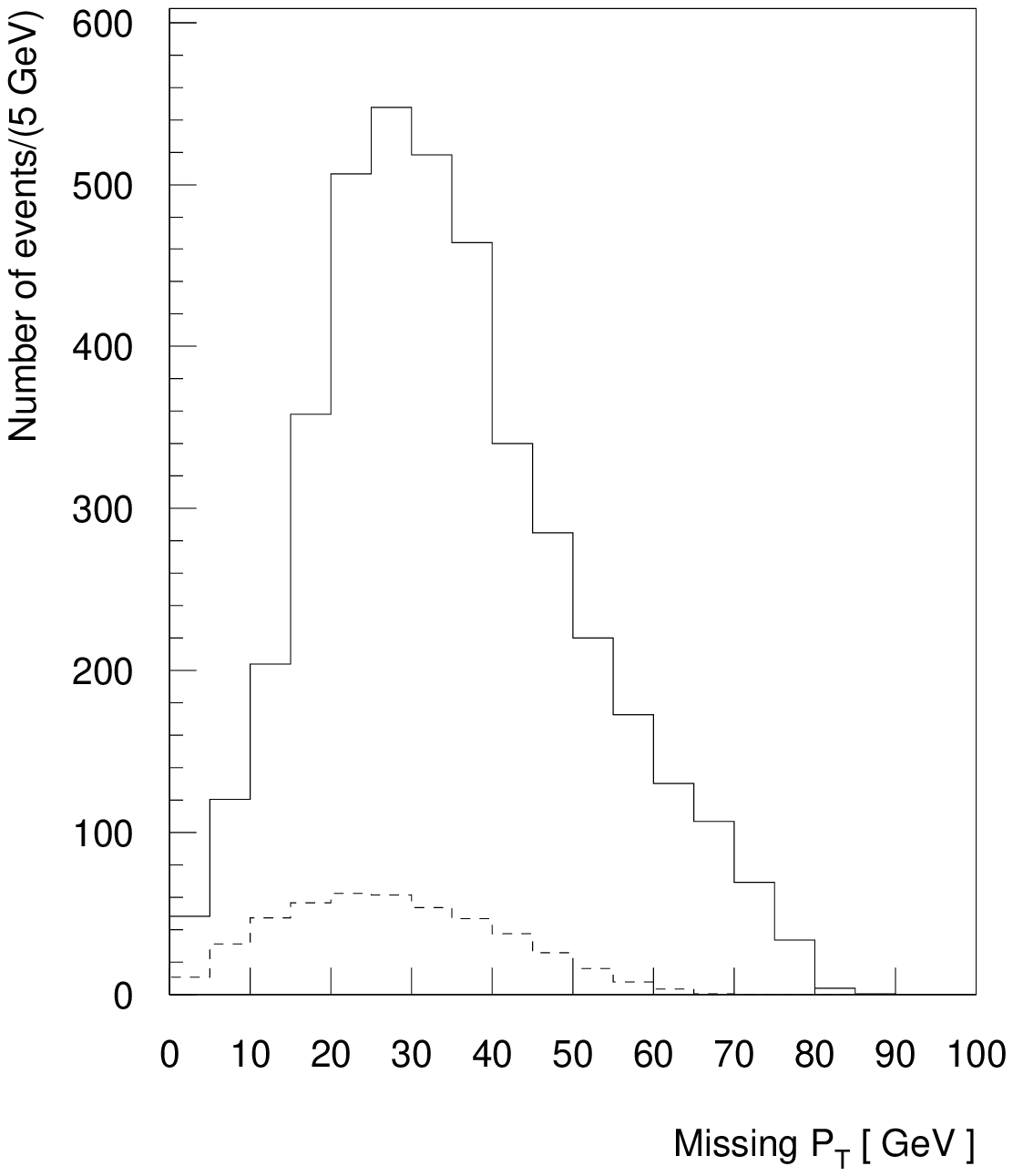}
    \caption{\label{mis-pt} The $\met$ distribution
     for the $lepton+2jets+\met$ signature for the background
	(solid line) and for the chargino pair production
     (dashed line).}
  \end{center}
\end{figure}

By taking into account the specific features  of the kinematic
variables for the signal and background shown above, the following set
of kinematic cuts has been worked out:
\begin{itemize} \item[1)]
several general cuts: ${E_T}_{lepton} > 10$~GeV, ${E_T}_{j1,2}  >
$10~GeV, isolation of the electron in terms of $\Delta
R=\sqrt{\Delta\phi_{ej}^2 + \Delta\eta_{ej}^2}> 0.3 $.
\item[2)]
missing transverse momentum  $> 10$~GeV, only a low cut, because
the $\met$ distributions for the signal and  background are similar, see
fig.~\ref{mis-pt}.
\item[3)] squared missing mass  $> 10000$~GeV$^2$
\item[4)] invariant electron-neutrino mass $>140$~GeV
\item[5)] di-jet mass $< 60$~GeV
\item[6)] $H_T<$~70~GeV
\end{itemize}

The consecutive action of these cuts is shown in Table~\ref{cuts}
for the number of events corresponding to the total integrated
luminosity 1000~pb$^{-1}$.  It is clearly seen that designed cuts shown
above suppress the background quite enough for the signal to be
subtracted. It should be pointed out that the upper  edge of the di-jet
mass distribution for the signal gives an important information about
the chargino and neutralino mass, it is equal to $(m_\chi^+
-m_\chi^0)$, so the cut on the di-jet mass has been chosen in a way it
does not affect the signal at all.  After all cuts have been applied, we
have 228 events for the signal and 15 events from the background which
is reduced by factor 0.002 from the parton level.

\begin{figure}[t]
  \vspace*{-1.0cm}
  \begin{center}
    \leavevmode
    \epsfxsize=12cm
    \epsffile{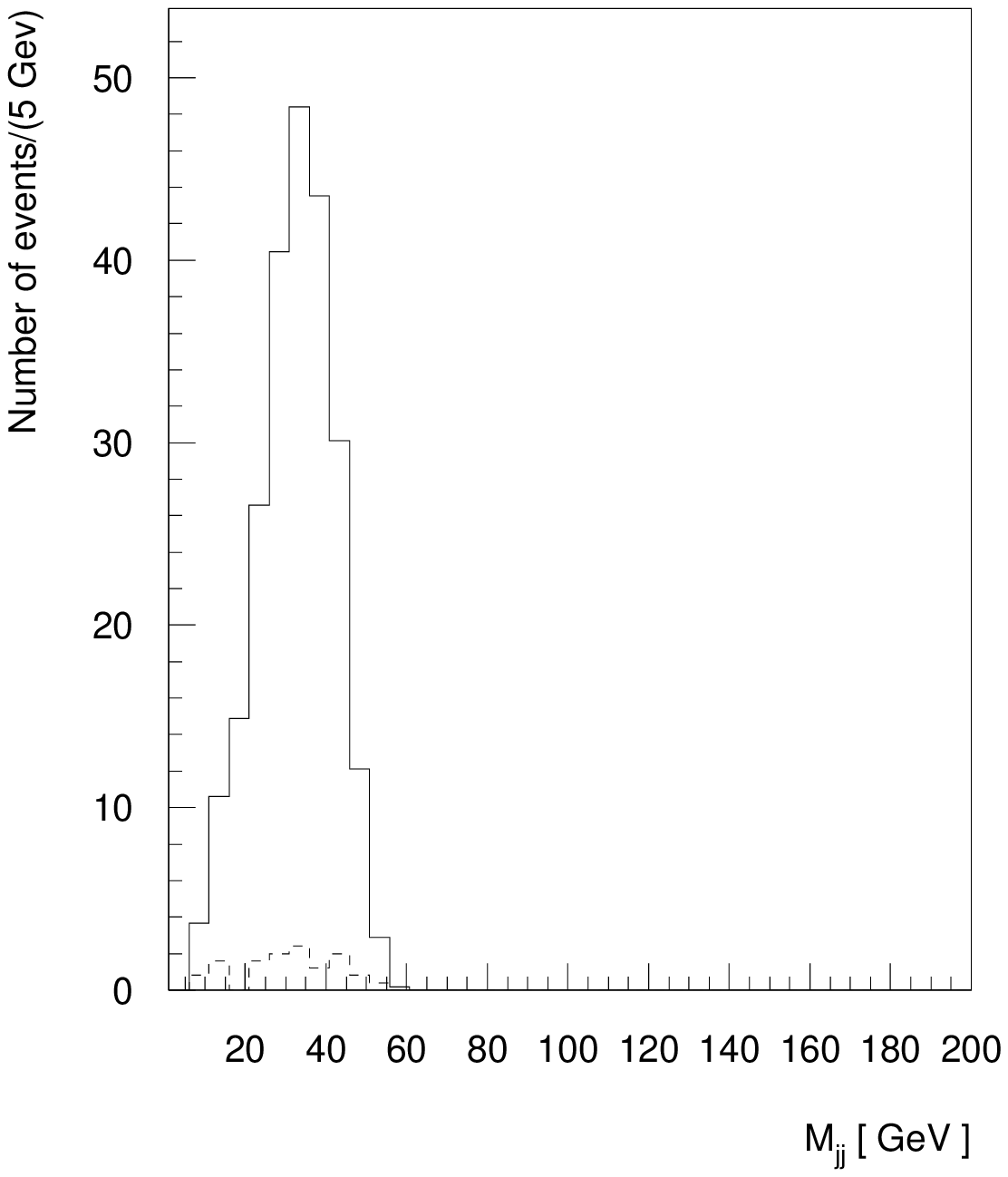}
    \caption{\label{di-jet-fin} The di-jet mass distribution
     for the $lepton+2jets+\met$ signature for the signal+background
	(solid line) and for the background (dashed line) after cuts
	have been applied.}
  \end{center}
\end{figure}

In fig.\ref{di-jet-fin} the di-jet mass distribution illustrating
the clear signal effect is shown after the whole set of cuts has been
applied.

\subsection{$4 \; jets + \met$ signature}

Let us turn now to the signature when both charginos decay hadronically
having the $4 \; jets + \met$ signature. For this case the signal ratio
with $BR=(0.67\cdot0.67)=0.45$ is 1.5 times higher than for the first
signature. So the cross section for this channel is
$2.9\cdot0.45=1.3$  pb.

The main SM backgrounds leading to the $4jets +\met$ signature are,
first of all, the $WW$ process when one $W$ decays hadronically and
the other decays into $\tau$-neutrino and $\tau$-lepton which decays
then hadronically giving an extra neutrino and
two jets. The branching ratio of this decay is
$2\cdot1/9[BR(W \to \tau \nu_\tau)]\cdot2/3[BR(W \to 2jets)]
\cdot0.65[BR(\tau \to \nu_\tau+hadrons)]=0.096$.
Thus the cross section of this process is
$22.9\cdot0.096=2.2$ pb.

We have also checked the other possible sources of backgrounds:
$e^+e^- \to \nu \overline\nu q \overline{q}$ with the consequent
gluon radiation of quarks,
(for example, the cross section of the
$e^- \to \nu_e \overline\nu_e u \overline{u}$ process
is equal to 0.038 pb, and after the di-jet radiation one can expect an
additional factor of ${\alpha_s}^2$,
and the total cross section of the processes
$e^+e^- \to \nu \overline\nu q \overline{q} \to \met + 4jets$
is of order 0.001~pb);
$e^+e^- \to \nu_e \overline\nu_e W^+ W^-$ process with the total
cross section $7.5\cdot10^{-7}$~pb and
$e^+e^- \to \nu_e \overline\nu_e Z^0 Z^0$ process
with the total cross section  $5.4\cdot10^{-9}$~pb. We can see that the
only real background is the first one mentioned above.

For the 1000~pb$^{-1}$ luminosity we can expect 1300 and 2200 events of
the signal and background, respectively, with the ratio  higher than
for the previous signature. At the same time  MC simulation shows
that it is more difficult to extract the signal for that type of events.
The missing mass distribution for the signal is softer than that for the
$lepton+2jets+\met$ signature because of the absence of the  neutrino,
while for the background it is wider and harder because of the presence
of additional neutrinos after the $\tau$-lepton decay.
Also, for the $4jet+\met$ signature there is no lepton in the final
state and we cannot use the invariant mass of a lepton and a neutrino as
a discriminator of the background.

There is the  reason for the fact that the signal ratio for
$4jet+\met$ becomes smaller than for $lepton+2jets+\met$ signature
after the detector simulation. It is related to the probability of the
jet to be reconstructed. For $WW$ events this probability is  higher
than that for chargino production with softer jets coming from
virtual $W$-boson with effective mass of 40~GeV which is half as large
as that for the real $W$-boson.  The estimated probability (for
parameters under the assumption) of the jet reconstruction from the real
$W$-boson is about 90\% while for the virtual $W$-boson  it is only
70\%.  It means that the signal with two or four jets  loses 50\% or
 75\% events respectively. This fact makes also the $lepton+2jets+\met$
signature a little bit more attractive for the chargino search.

At the same time the $4-jet+\met$  channel  is complementary  to the
$l+2jet+\met$ one and can give information about the branching ratios
of chargino decay; it would be  an independent confirmation
of the possible chargino existence.

\begin{figure}[t]
  \vspace*{-1.0cm}
  \begin{center}
    \leavevmode
    \epsfxsize=12cm
    \epsffile{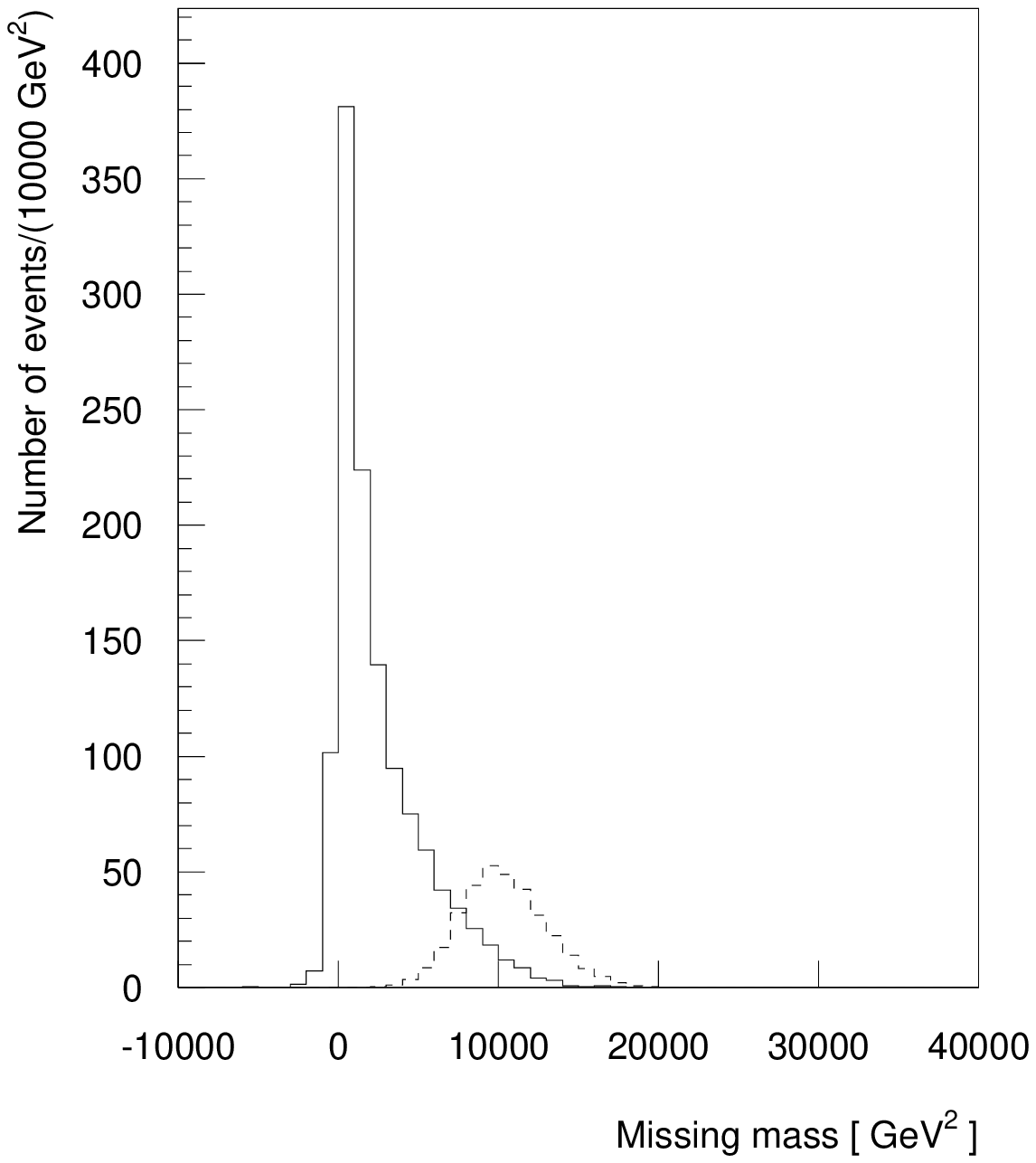}
    \caption{\label{miss-mass-jj} The missing mass distribution
     for the $4jets+\met$ signature for the background (solid line)
	and for the chargino pair production (dashed line).}
  \end{center}
\end{figure}

\begin{figure}[t]
  \vspace*{-1.0cm}
  \begin{center}
    \leavevmode
    \epsfxsize=12cm
    \epsffile{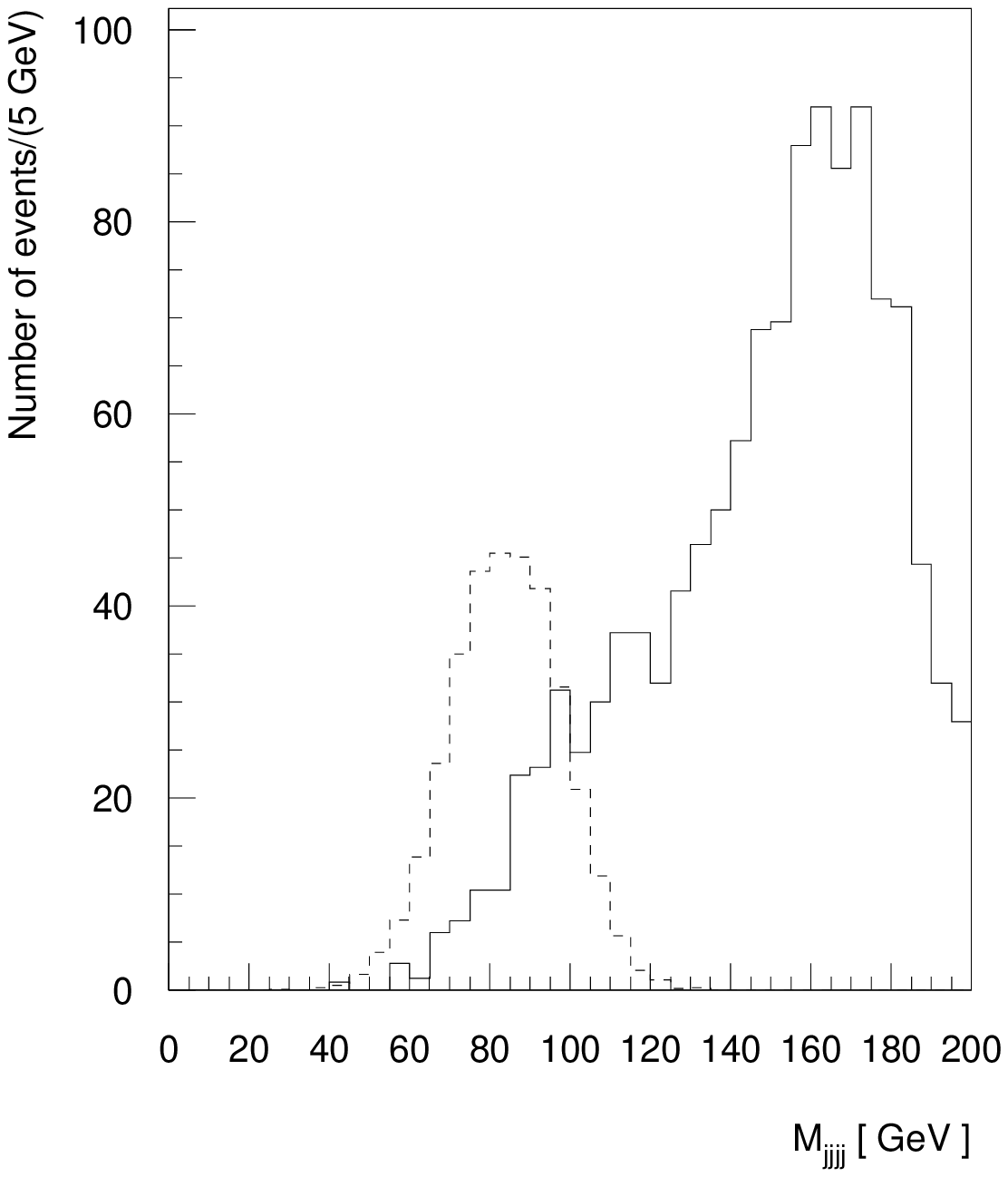}
    \caption{\label{dijet-mass-jj} The di-jet mass distribution
     for the $4jets+\met$ signature for the background (solid line)
	and for the chargino pair production (dashed line).}
  \end{center}
\end{figure}

\begin{figure}[t]
  \vspace*{-1.0cm}
  \begin{center}
    \leavevmode
    \epsfxsize=12cm
    \epsffile{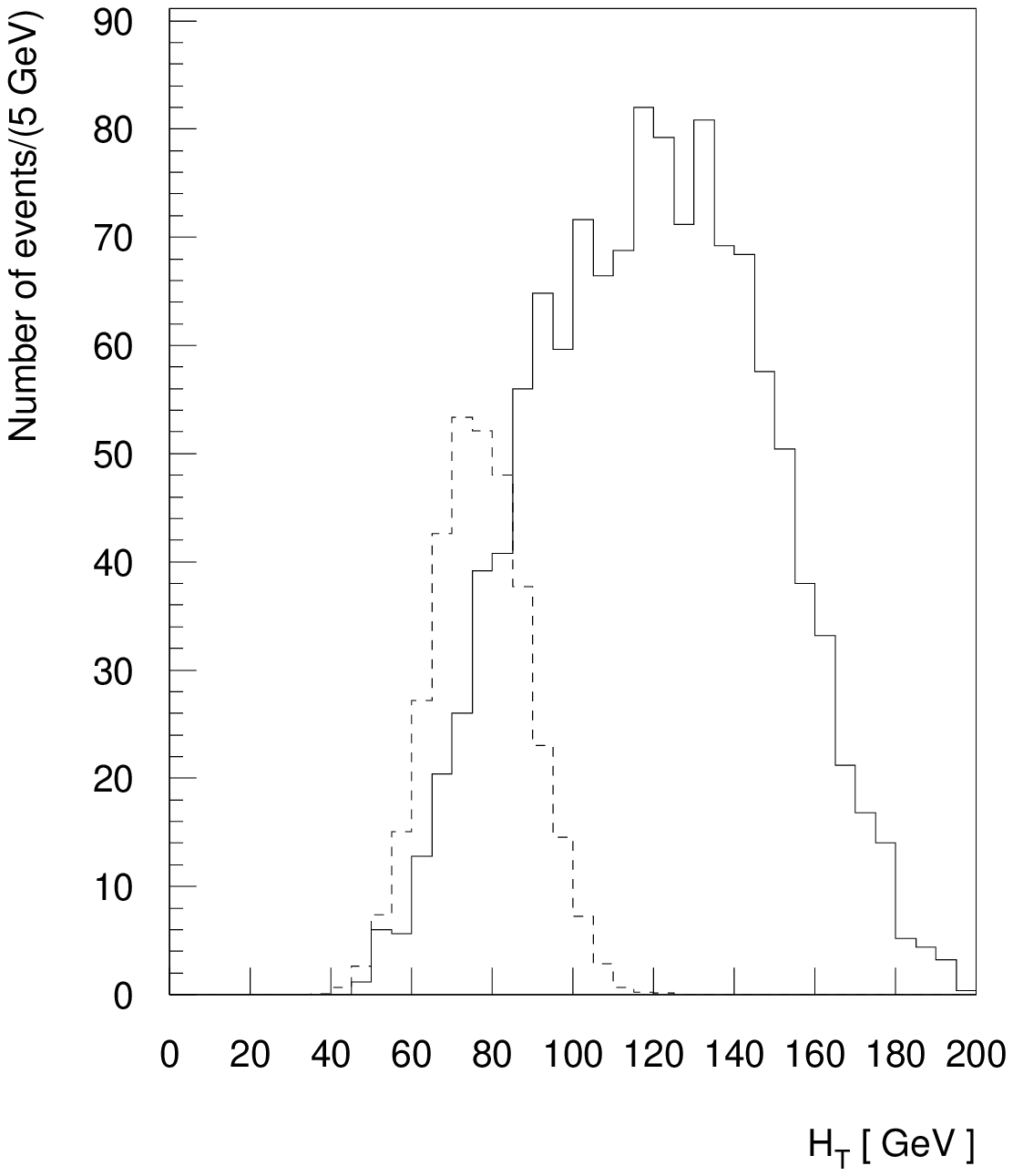}
    \caption{\label{ht-jj} The $H_T$ distribution
     for the $4jets+\met$ signature for the background (solid line)
	and for the chargino pair production (dashed line).}
  \end{center}
\end{figure}

\begin{figure}[t]
  \vspace*{-1.0cm}
  \begin{center}
    \leavevmode
    \epsfxsize=12cm
    \epsffile{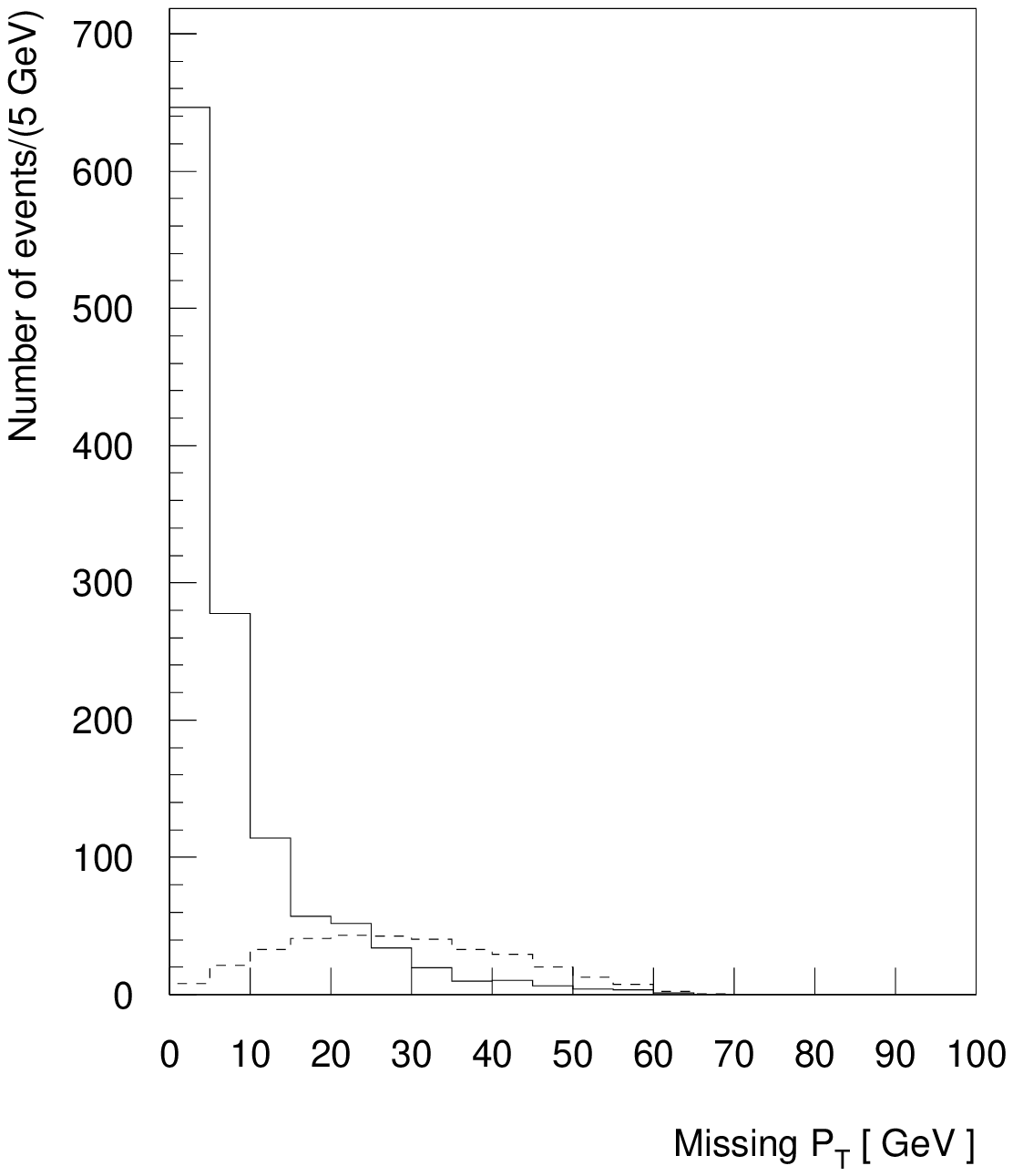}
    \caption{\label{mis-pt-jj} The $\met$ distribution
     for the $4jets+\met$ signature for the background (solid line)
	and for the chargino pair production (dashed line).}
  \end{center}
\end{figure}

For the background reduction we chose kinematical variables similar
to those for the previous signature:
\begin{itemize}
\item[1)] Squared missing mass.
For the background it has a peak around zero, but this distribution,
as it was told above, is wider than that for the $l+2jet+\met$  events
and softer for the signal, see fig.~\ref{miss-mass-jj}.
\item[2)] Invariant four-jet mass, which for the $WW$-background is
concentrated around $2M_W$, while for the signal it has a peak around
2$m_{\chi^+}-m_{\chi^0}$, see fig.~\ref{dijet-mass-jj}.
\item[3)] $H_T$ variable, see fig.~\ref{ht-jj}.
\end{itemize}
The following set of kinematic cuts has been designed for this
signature:
\begin{itemize}
\item[1)] several general cuts:  ${E_T}_{j1,2,3,4}>$ 10~GeV,
\item[2)] missing transverse momenta  $>15$~GeV,
to reject a big amount of events which have nothing to do with
the $\tau$-lepton in the final state. Usually these events are from
the 4-jet  $WW$-decay with small missing transverse momenta,
see fig.~\ref{mis-pt-jj}.
\item[3)] squared missing mass  $> 5000$ GeV$^2$
\item[4)] four-jet mass $< 120$ GeV
\item[5)] $H_T<$ 100 GeV
\end{itemize}

The consecutive action of these cuts is shown in Table~\ref{cuts-jj}
for the number of events corresponding to the same  integrated
luminosity as before.  It is also clearly seen that the designed cuts
suppress the background quite enough for the signal to be
subtracted.

\begin{figure}[t]
  \vspace*{-1.0cm}
  \begin{center}
    \leavevmode
    \epsfxsize=12cm
    \epsffile{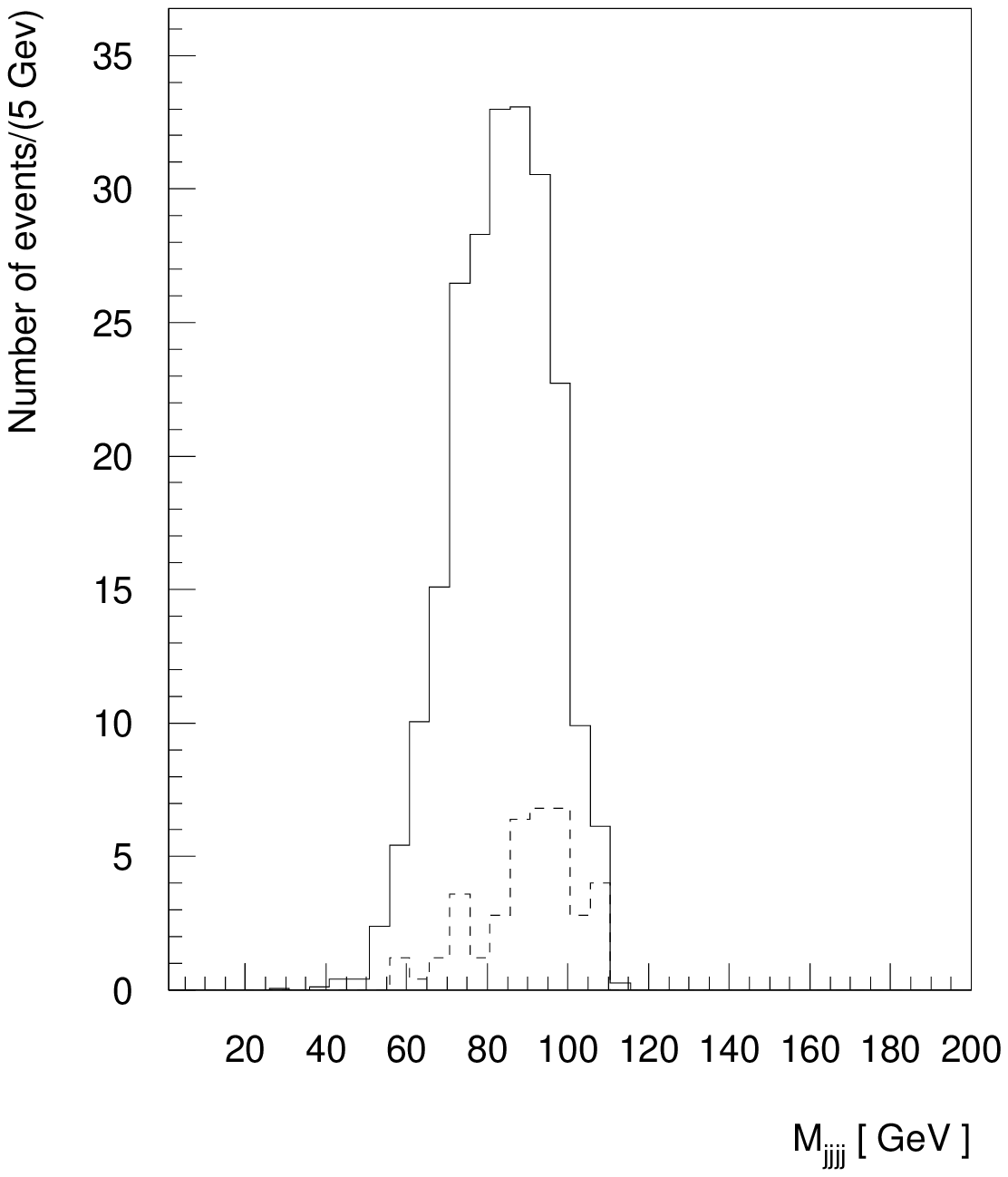}
    \caption{\label{di-jet-finjj} The di-jet mass distribution
     for the $4jets+\met$ signature for the signal+background
	(solid line) and for the background (dashed line) after cuts
	have been applied.}
  \end{center}
\end{figure}

After all cuts have been applied, we have 261 events for the signal
and 43 from the background which is reduced by factor 0.02.
We can see that the reduction factor for the background for this
signature is 10 times smaller than that for the $l+2jets+\met$
signature for  the reasons explained above.
But nevertheless the cuts for the "jet" signature have a big affect
on the background. In fig.\ref{di-jet-finjj} the four-jet mass
distribution, with clear signal effect similar to the previous
signature, is shown after the whole set of cuts
has been applied.

The upper edges of the four-jet mass distribution determine
the $2\cdot(m_\chi^+ -m_\chi^0)$
value which should be consistent with the value $(m_\chi^+ -m_\chi^0)$
coming from the $l+2jets+\met$ signature study and it is complementary,
which can improve determination of the chargino and neutralino masses.
The procedure of extracting information about the chargino and
neutralino from the endpoints of $M_{jj}$ and $E_{jj}$ has been
described in~\cite{feng}. The same procedure should be used in the case
of the endpoints for $M_{jjjj}$ and $E_{jjjj}$.  The accuracy of the
determination of the chargino and neutralino mass is expected to be of
order 5~GeV.

\section{Conclusions}

In this paper, we have investigated the potential of the LEP II collider
for the search of the chargino signal. The charged fermion,
light chargino, is the most preferable SUSY particle to be discovered
the first at LEP II.

The total production rate of pair chargino production has been
calculated. In the mass range within the reach of LEP II, the cross
section of chargino production is of order 1~pb.

The decay channels of the chargino have  been studied. The
only possibility of three-body decay is shown. The decay width as well
as branching ratios have been calculated. It is shown that decay of the
chargino into the next-to-lightest neutralino (which opens the cascade
decay of the chargino) is suppressed enough for not to be taken  into
account. The most preferable signatures for chargino search, namely
$lepton+2jets+\met$ and $4jets+\met$ have been studied. The signature
with  four jets and missing transverse momenta has not been
investigated in  details before.

For analysing of the signal and background kinematics in order to
suppress the latter, the MC generator has been created. In comparison
with the previous papers related to the chargino search at LEP II,
not only smearing but also the effects of hadronization and jet
reconstruction as well as final state  radiation have been taken into
account, which is especially important for the $4-jets+\met$ signature.
Based on the study of the signal and
background  kinematics,  the set of cuts for the signal extraction
has been designed. These cuts suppress the background quite enough
for the signal to be clearly seen. The information about chargino
and neutralino masses can be extracted from the endpoints
of the $M_{jj}$ and $E_{jj}$ or $M_{jjjj}$ and $E_{jjjj}$
distributions with an accuracy of order 5~GeV.

\begin{figure}[t]
  \vspace*{-1.0cm}
  \begin{center}
    \leavevmode
    \epsfxsize=12cm
    \epsffile{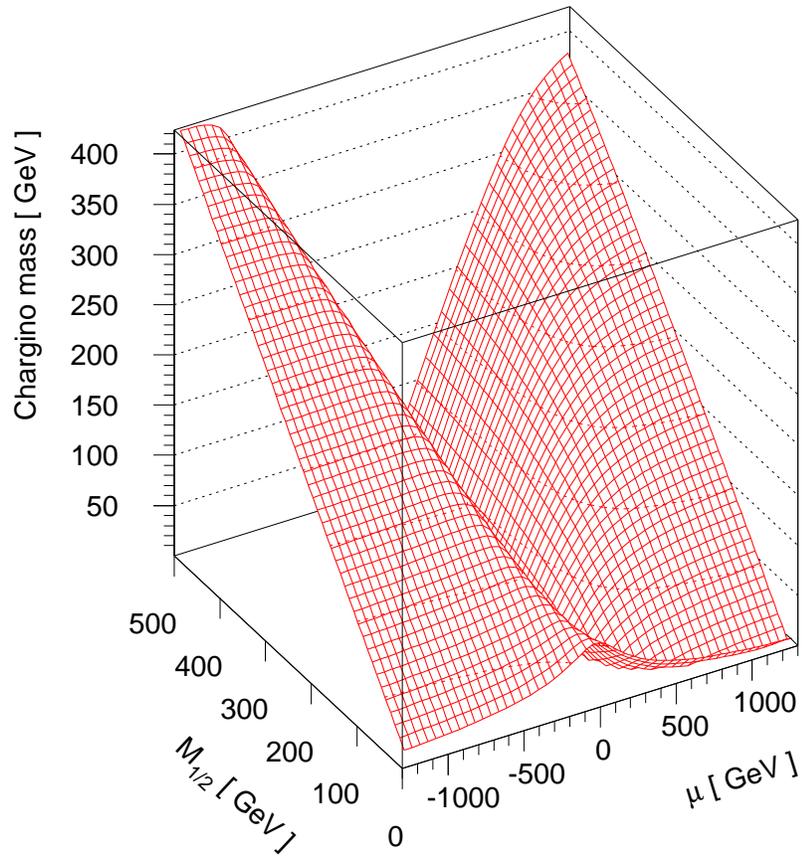}
    \caption{\label{2d} The chargino mass as a function of $m_{1/2}$
    and $\mu$ for low $\tan\beta$}
  \end{center}
\end{figure}

\begin{figure}[t]
  \vspace*{-1.0cm}
  \begin{center}
    \leavevmode
    \epsfxsize=12cm
    \epsffile{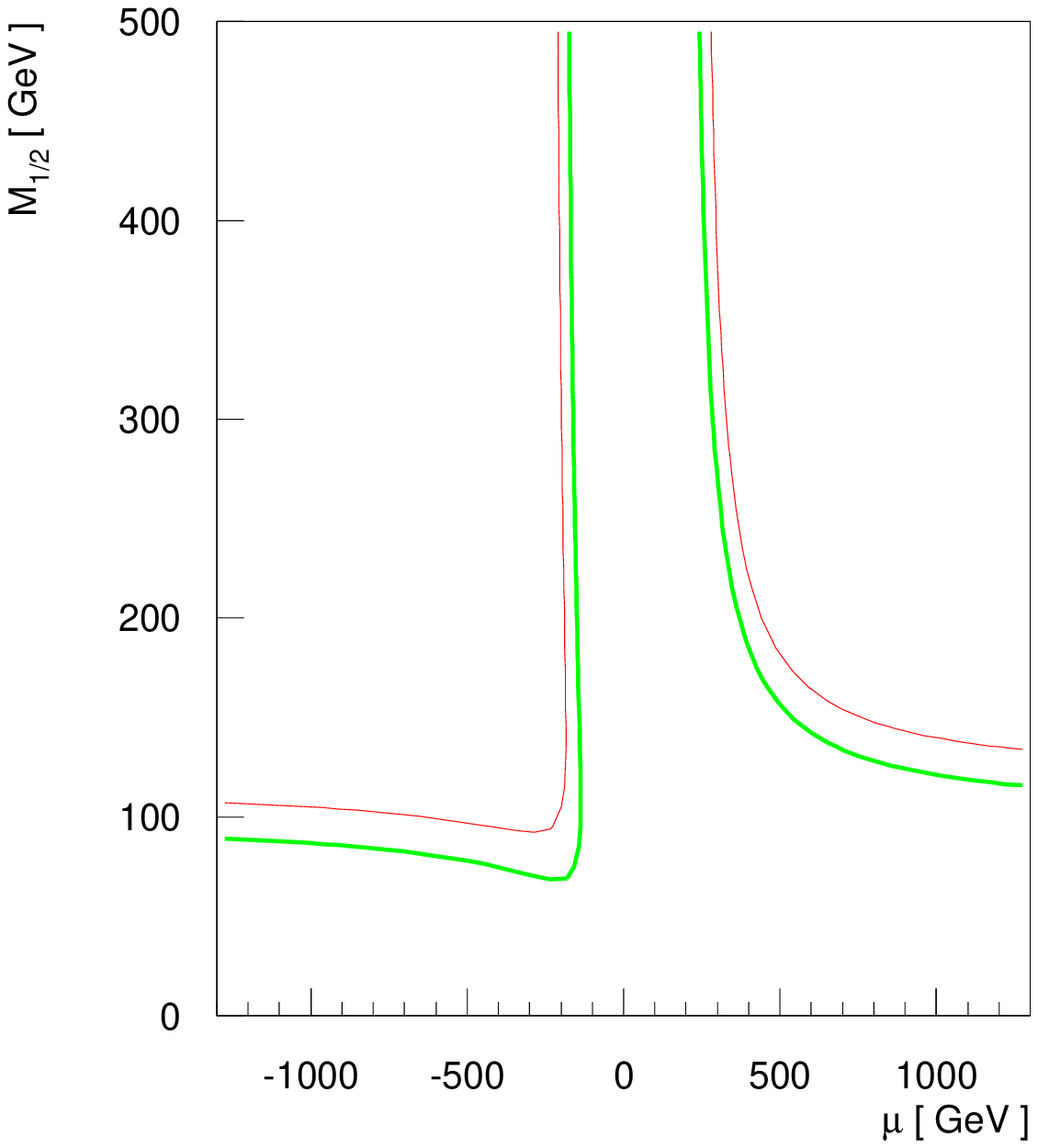}
        \put(-130,100){1}
        \put(-240,129){2}
        \put( -50,149){2}
    \caption{\label{limits} The region in the $m_{1/2}-\mu$ plane
     for low $\tan\beta$ excluded for $m_{\chi^\pm}>85$~GeV (1) and
	can be excluded for LEP II $m_{\chi^\pm}>99$~GeV (2).}
  \end{center}
\end{figure}

\begin{figure}[t]
  \vspace*{-1.0cm}
  \begin{center}
    \leavevmode
    \epsfxsize=12cm
    \epsffile{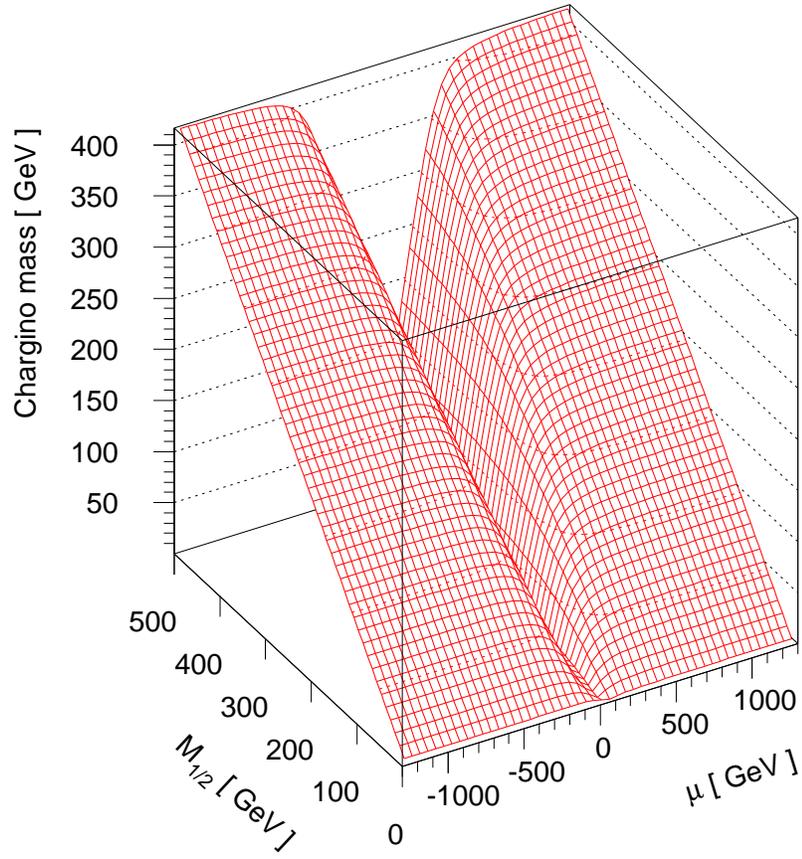}
    \caption{\label{2d-htb} The chargino mass as a function of
	$m_{1/2}$ and $\mu$ for the high $\tan\beta$ scenario.}
  \end{center}
\end{figure}

\begin{figure}[t]
  \vspace*{-1.0cm}
  \begin{center}
    \leavevmode
    \epsfxsize=12cm
    \epsffile{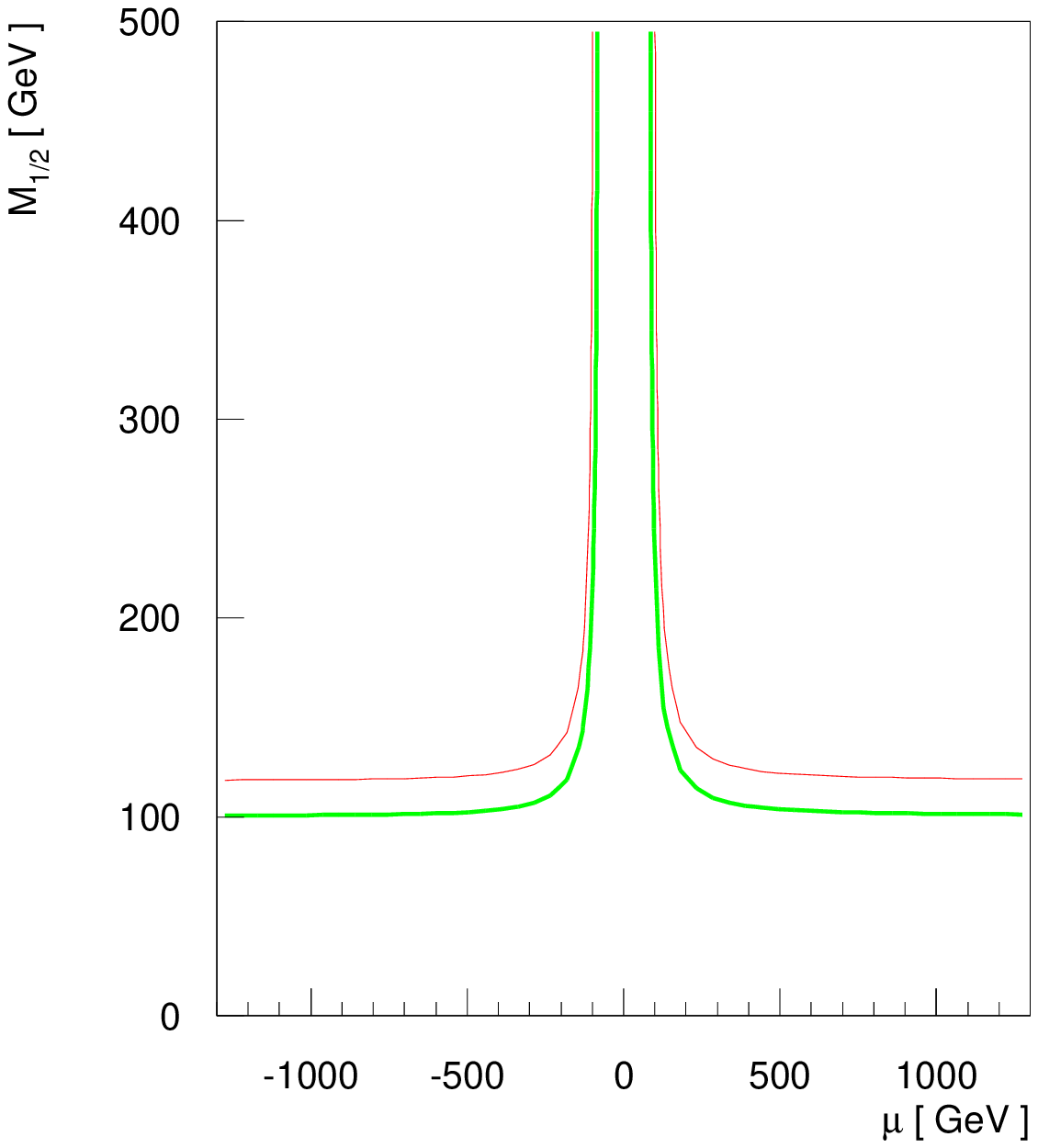}
        \put(-150,120){1}
        \put(-230,139){2}
        \put( -70,139){2}
    \caption{\label{limits-htb} The region in the $m_{1/2}-\mu$ plane
     for high $\tan\beta$ excluded for $m_{\chi^\pm}>85$~GeV (1) and
     can be excluded for LEP II $m_{\chi^\pm}>99$~GeV (2).}
  \end{center}
\end{figure}

The limits on the chargino mass which could be obtained at LEP II is
very close (1-2~GeV) to the kinematic limit of the machine.  The
chargino discovery  would shed light on the supersymmetry parameter
space especially on $m_{1/2}$ and $\mu$ which are directly related to
the chargino and neutralino masses.  One can draw the chargino mass as
a 2-D function of the $m_{1/2}$ and $\mu$ parameters (fig.~\ref{2d} for
low  $\tan\beta$ and fig.~\ref{2d-htb} for  high $\tan\beta$ scenarios)
and study  the limits of $m_{1/2}$ and  $\mu$  for a fixed chargino
mass which will be obtained at LEP II. In fig.~\ref{limits} (low
tan$\beta$) and  fig.~\ref{limits-htb} (high tan$\beta$) two
regions in the $m_{1/2}$ and $\mu$ plane are shown excluding SUSY
parameters for the 85~GeV (the present limit) and
99~GeV (the limit for $\sqrt{s}=200$ GeV) chargino. For example, for
typical values of $\mu=$500 and 1000~GeV the limit on $m_{1/2}$ can be
extended for the low $\tan\beta$ scenario from 75 up to 100~GeV and
from 80 to 107~GeV, respectively, for LEP II with $\sqrt{s}=200$ GeV.
For high $\tan\beta$ the respective limits will be extended from 101 to
118~GeV for $\mu =500$ GeV and from 102 to 121~GeV for $\mu =100$ GeV.


\subsection*{Acknowledgments}

We would like to thank
D.I. Kazakov and  W. de Boer for valuable discussions and comments.
We are also grateful to all members of the CompHEP group
for the help in adapting CompHEP for the SUSY model and useful
discussions, especially, to E.Boos,  V.Ilyin and  A.Pukhov.

The financial support of the Russian Foundation for Basic
Research  (grants \# 96-02-17379-a,
\#96-02-19773-a)  and ICFPM in 1996 is acknowledged.

\vspace{2cm}
\section*{Figures}
%
%

\clearpage

%
%

\clearpage

%
%

\clearpage

%
%

\clearpage

%
%

\clearpage

%
%

\clearpage

%
%
\begin{figure}[t]
  \vspace*{-1.0cm}
  \begin{center}
    \leavevmode
    \epsfxsize=12cm
    \epsffile{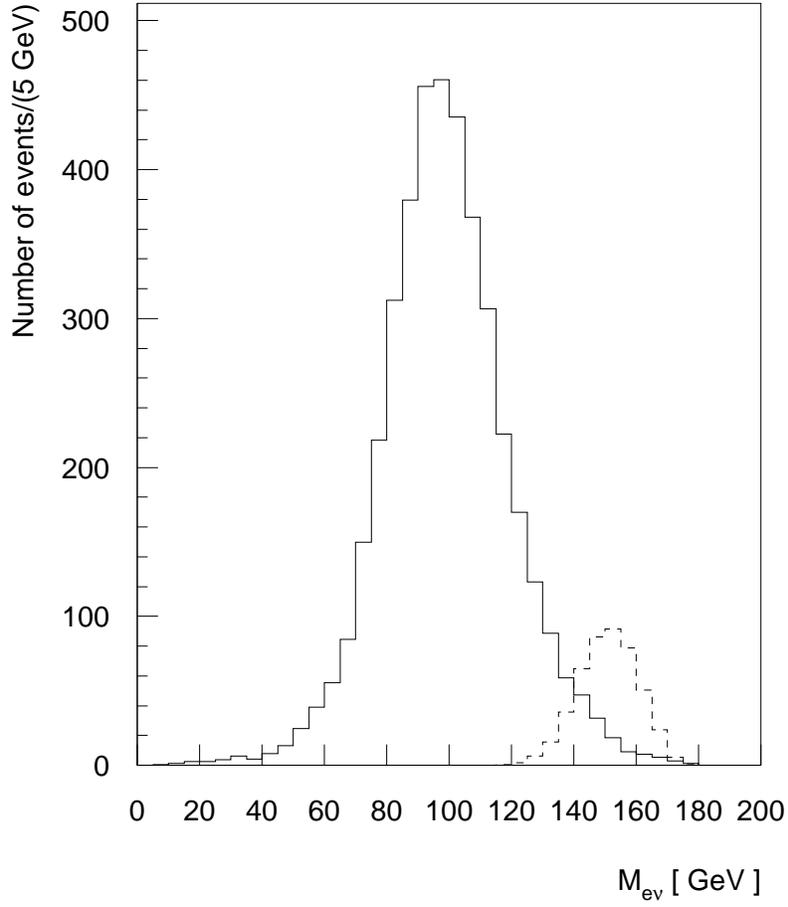}
    \caption{\label{men} The invariant electron-neutrino mass
     distribution for the $lepton+2jets+\met$ signature for the
     background (solid line) and for the chargino pair production
	(dashed line).}
   \end{center}
\end{figure}

\clearpage

%
%

\clearpage

%
%

\clearpage

%
%

\clearpage

%
%

\clearpage

%
%

\clearpage

%
%

\clearpage

%
%

\clearpage

%
%

\clearpage

%
%

\clearpage

%
%

\clearpage

%
%

\clearpage

%
%

\clearpage

%
%

\clearpage

\section*{Tables}

\vspace{1cm}

\begin{table}[h]
\begin{center}
\begin{tabular}{|c|r|r|}
\hline
 & $\tan\beta=1.52$ & $\tan\beta=41.2$ \\ \hline
$m_0$ & 400 GeV & 800 GeV \\ \hline
$m_{1/2}$ & 111 GeV & 88 GeV \\ \hline
$\mu(0)$ & 1260 GeV & --270 GeV \\ \hline
$A(0)$ & 0 GeV & 1256 GeV \\ \hline
\end{tabular}
\caption{\label{fit}
The "best fit" SUSY parameters for the low and high $\tan\beta$
scenarios.}
\end{center}
\end{table}
\begin{table}[h]
\begin{center}
\begin{tabular}{|c|r|r|}
\hline
& $\tan\beta=1.52$ & $\tan\beta=41.2$ \\ \hline
$\tilde\chi^\pm_1$ &   82 GeV &  70 GeV  \\ \hline
$\tilde\chi^\pm_2$ &  549 GeV & 304 GeV  \\ \hline
$\tilde\chi^0_1$   &   41 GeV &  35 GeV  \\ \hline
$\tilde\chi^0_2$   &   83 GeV &  69 GeV  \\ \hline
$\tilde\chi^0_{3,4}$   & $\approx$ 540 GeV & $\approx$295 GeV  \\ \hline
$\tilde t_1$       &  140 GeV & 504 GeV  \\ \hline
$\tilde b_1$       &  383 GeV & 675 GeV  \\ \hline
$\tilde\nu_\tau$   &  407 GeV & 818 GeV  \\ \hline
\end{tabular}
\caption{\label{mass}
The mass spectrum of some SUSY particles for the
"best fit"  for the low and high $\tan\beta$ scenarios.}
\end{center}
\end{table}
\begin{table}[h]
\begin{center}
$
\begin{array}{|r|l|l|r|}
\hline
  & Process & \Gamma (GeV)  & BR (\%) \\ \hline
1 & \chi^+ \to c\bar{b}\chi^0         & 0.83921 \cdot 10^{-7}
	&  0.0\% \\ \hline
2 & \chi^+ \to u\bar{b}\chi^0         & 0.69445 \cdot 10^{-9}
	&  0.0\% \\ \hline
3 & \chi^+ \to c\bar{s}\chi^0         & 0.45610 \cdot 10^{-4}
	& 31.3\% \\ \hline
4 & \chi^+ \to u\bar{s}\chi^0         & 0.23505 \cdot 10^{-5}
	&  1.6\% \\ \hline
5 & \chi^+ \to c\bar{d}\chi^0         & 0.23345 \cdot 10^{-5}
	&  1.6\% \\ \hline
6 & \chi^+ \to u\bar{d}\chi^0         & 0.45991 \cdot 10^{-4}
	& 31.3\% \\ \hline
7 & \chi^+ \to \tau^+ \nu_\tau \chi^0 & 0.16671 \cdot 10^{-4}
	& 11.4\% \\ \hline
8 & \chi^+ \to \mu^+ \nu_\mu \chi^0   & 0.16303 \cdot 10^{-4}
	& 11.4\% \\ \hline
9 & \chi^+ \to e^+ \nu_e \chi^0       & 0.16303 \cdot 10^{-4}
	& 11.4\% \\ \hline
\end{array}
$
\end{center}
\caption{\label{width} Chargino decay widths and branching ratios.}
\end{table}
\begin{table}[h]
\begin{center}
\begin{tabular}{|r|r|r|r|r|}
\hline
Cut & $WW:l+j+\met$ & $signal:l+j+\met$ \\ \hline
0 & 6600 & 900 \\ \hline
1 & 4385 & 346 \\ \hline
2 & 4206 & 317 \\ \hline
3 &   31 & 301 \\ \hline
4 &   26 & 268 \\ \hline
5 &   26 & 268 \\ \hline
6 &   15 & 228 \\ \hline
\end{tabular}
\caption{\label{cuts} Numbers of events for the signal and background
for the $l+2jets+\met$ signature
for the consecutive cut application for the integrated luminosity
1000 pb$^{-1}$.}
\end{center}
\end{table}
\begin{table}[h]
\begin{center}
\begin{tabular}{|r|r|r|r|r|}
\hline
Cut  & $WW: 4jet+\met$ & $signal:4jet+\met $\\ \hline
0 & 2200 & 1300 \\ \hline
1& 1183 & 261 \\ \hline
2 &  176 & 212 \\ \hline
3 &   44 & 190  \\ \hline
4 &   44 & 190 \\ \hline
5 &   43 & 187 \\ \hline
\end{tabular}
\caption{\label{cuts-jj} Number of events for the signal and background
for the $4jets+\met$ signature
for the consecutive cut application for the integrated luminosity
1000 pb$^{-1}$.}
\end{center}
\end{table}

\end{document}